\documentclass[]{article}
\usepackage{emulateapj}
\usepackage{graphicx}
\usepackage{multirow}
\usepackage{amsmath}
\usepackage{pdflscape}
\usepackage{hyperref}
\usepackage{amsmath}

\advance \voffset by -0.5cm\relax

\usepackage{color}

\def\msun{{\rm ~M}_{\odot}}
\def\rsun{{\rm ~R}_{\odot}}
\def\myr{{\rm ~Myr}}

\def\zsun{{\rm ~Z}_{\odot}}
\def\zsunn{{\rm Z}_{\odot}}
\def\kms{{\rm ~km} {\rm ~s}^{-1}}
\def\mpy{{\rm ~Mpc}^{-3} {\rm ~yr}^{-1}}
\def\msy{{\rm ~M}_{\odot} {\rm ~yr}^{-1}}

\begin{document}

\title{Compact Binary Merger Rates: Comparison with LIGO/Virgo Upper Limits}

 \author{Krzysztof Belczynski\altaffilmark{1},
         Serena Repetto\altaffilmark{2},
         Daniel E. Holz\altaffilmark{3}, 
         Richard O'Shaughnessy\altaffilmark{4},
         Tomasz Bulik\altaffilmark{1},          
         Emanuele Berti\altaffilmark{5,6},
         Christopher Fryer\altaffilmark{7}, 
         Michal Dominik\altaffilmark{1}, 
}

 \affil{
     $^{1}$ Astronomical Observatory, Warsaw University, Al.
            Ujazdowskie 4, 00-478 Warsaw, Poland\\
     $^{2}$ Department of Astrophysics/IMAPP, Radboud University Nijmegen, PO Box 9010, 
            6500 GL Nijmegen, The Netherlands\\
     $^{3}$ Enrico Fermi Institute, Department of Physics, and Kavli Institute
            for Cosmological Physics, University of Chicago, Chicago, IL 60637 \\
     $^{4}$ Center for Computational Relativity and Gravitation, Rochester
            Institute of Technology, Rochester, New York 14623, USA \\ 
     $^{5}$ Department of Physics and Astronomy, The University of Mississippi,
            University, MS 38677, USA \\  
     $^{6}$ CENTRA, Departamento de F\'isica, Instituto Superior T\'ecnico, Universidade 
            de Lisboa, Avenida Rovisco Pais 1, 1049 Lisboa, Portugal \\
     $^{7}$ CCS-2, MSD409, Los Alamos National Laboratory, Los Alamos, NM 87545 \\
}

\begin{abstract}
We compare evolutionary predictions of double compact object merger rate 
densities with initial and forthcoming LIGO/Virgo upper limits. We find that: 
{\em (i)} Due to the cosmological reach of advanced detectors, current conversion 
methods of population synthesis predictions into merger rate densities are 
insufficient. 
{\em (ii)} Our optimistic models are a factor of $18$ below the initial LIGO/Virgo 
upper limits for BH-BH systems, indicating that a modest increase in observational 
sensitivity (by a factor of $\sim 2.5$) may bring the first detections or first 
gravitational wave constraints on binary evolution. 
{\em (iii)} Stellar-origin massive BH-BH mergers should dominate event rates in 
advanced LIGO/Virgo and can be detected out to redshift $z\simeq2$ with templates 
including inspiral, merger, and ringdown. Normal stars ($<150\msun$) can produce 
such mergers with total redshifted mass up to $M_{\rm tot,z} \simeq 400\msun$. 
{\em (iv)} High black hole natal kicks can severely limit the formation of massive 
BH-BH systems (both in isolated binary and in dynamical dense cluster evolution), 
and thus would eliminate detection of these systems even at full advanced LIGO/Virgo 
sensitivity. We find that low and high black hole natal kicks are allowed by current 
observational electromagnetic constraints.
{\em (v)} The majority of our models yield detections of all types of mergers 
(NS-NS, BH-NS, BH-BH) with advanced detectors. Numerous massive BH-BH merger 
detections will indicate small (if any) natal kicks for massive BHs. 
\end{abstract}


\section{Introduction}

Most massive stars are found in binary systems (e.g., Garcia \& Mermilliod 2001; 
Kiminki et al. 2007; Kobulnicky \& Fryer 2007; Sana et al. 2012; Kobulnicky et al. 2014). 
During the evolution of these stars the binaries can experience component merger 
during common envelope (CE) phases (e.g., Webbink 1984) or disruption during supernova 
(SN) explosions (e.g., Tauris \& Takens 1998) in which individual stars form neutron 
stars (NSs) or black holes (BHs). The massive binaries which survive these processes 
form double compact objects: NS-NS, BH-BH, or mixed BH-NS systems (e.g., Belczynski, 
Kalogera \& Bulik 2002). These remnant systems are subsequently subject to angular 
momentum loss via the emission of gravitational waves (GWs) and their orbital separation 
decreases (Peters \& Mathews 1963; Weisberg \& Taylor 2005). Finally, the two compact 
objects merge into a single compact object giving rise to a strong GW signal (Einstein 
1918). 

The LIGO/Virgo network of ground-based interferometric observatories has been 
designed to search for gravitational-waves, including those resulting from the
merger of compact binary systems~(Abbott et al.~2009, Accadia et al. 
2012)\footnote{\url{http://www.ligo.caltech.edu/}; \url{http://www.virgo.infn.it/}}.
Theoretical predictions for near-future detection probabilities were compiled and 
presented by Abadie et al. (2010). Initial LIGO/Virgo observations were
concluded in $2010$ without the detection of a GW signal (e.g., Abadie et al. 2012). 
The instruments are currently being upgraded and the network is resuming its 
operation (2015) and will reach target sensitivity by $\sim 2019$. 

It has long been recognized that double compact object binaries are likely to be
important sources for gravitational-wave observatories (Thorne 1987, Schutz
1989). One of the most promising sources is the inspiral and merger of NS-NS
systems. There are several known Galactic double neutron star binaries with 
merger times shorter than the Hubble time (e.g., Kim, Kalogera \& Lorimer 2010).
Moreover, observational evidence points to mergers of neutron stars with other 
neutron stars or black holes as progenitors of short gamma ray bursts (GRBs, 
Berger 2013). Recently a first candidate for a kilonova, which is expected to accompany  
short GRBs if they originate from NS-NS or BH-NS mergers, has been observed (Berger, 
Fong \& Chornock 2013; Tanvir et al. 2013). Surprisingly, a second candidate
has been found to accompany a long ($\sim 100$ s) GRB (Yang et al. 2015).
Double black hole binaries, on the other hand, remain undetected. Flanagan \& 
Hughes (1998) emphasized that BH-BH binaries are particularly suitable for
gravitational-wave detection, since 
their signals are quite strong at frequencies where ground-based GW
observatories have excellent sensitivity, making them detectable to great 
distances. Recent theoretical predictions indicate that these systems may either 
dominate gravitational-wave observations (e.g., Belczynski et al. 2010a) or be 
totally absent in the local Universe (e.g., Mennekens \& Vanbeveren 2014). 

In this study we compare merger rates from our evolutionary calculations of double 
compact object binaries (Dominik et al. 2012, 2013, 2015) with the latest LIGO/Virgo 
upper limits (Abadie et al. 2012; Aasi et al. 2013a, 2014a, 2014b). We also
compare our predictions with the expected upper limits of double compact object  
mergers from advanced GW instruments (Harry et al. 2010; Acernese et al. 2015; Abadie 
et al. 2015). 
Our synthetic models cover a wide mass range, from light NS-NS binaries with 
total mass as low as $\gtrsim 2 \msun$ to massive stellar-origin BH-BH mergers with 
total mass as high as $140\msun$. For comparison, the low-mass LIGO/Virgo search 
was performed for a total mass range $2$--$25\msun$ (Abadie et al. 2012), the 
high-mass search was for $25$--$100\msun$ (Aasi et al. 2013a), 
and the very high-mass search was for $100$--$450\msun$ 
(Aasi et al. 2014b). 

We put special emphasis on BH natal kicks and their potential effect on BH-BH 
merger rates. This is especially timely as several recent studies on BH natal 
kicks have recently appeared (e.g., Wong et al. 2014; Reid et al. 2014; Repetto 
\& Nelemnas 2015; Mandel 2016).  
In the past we have studied uncertainties associated with common envelope evolution  
(e.g., Belczynski et al. 2007), as well as the impact of metallicity on our predictions 
(e.g., Belczynski et al. 2010a). 

An overview of our paper is as follows: 
In Sec.~\ref{calcul} we describe our merger rate density calculations, while 
Sec.~\ref{models} gives details of our evolutionary modeling. 
In Sec.~\ref{obsligo} we present the existing initial LIGO/Virgo and forthcoming 
advanced LIGO/Virgo upper limits. 
Our main result, the comparison of population synthesis models with LIGO/Virgo
upper limits, is given in Sec.~\ref{results}.
A detailed discussion of BH natal kicks is included in Sec.~\ref{BHkicks}, while
BH-BH and NS-NS merger rates are discussed in Secs.~\ref{nobhbh} and~\ref{nsns},
respectively. Finally, we list our conclusions in Sec.~\ref{conc}.

\section{Merger rate density estimates}
\label{calcul}
 
We have employed a set of publicly available evolutionary models from the Synthetic 
Universe database (\url{http://www.syntheticuniverse.org}) that provide physical 
properties and merger rates of NS-NS, BH-NS and BH-BH binaries. The calculations of 
the mergers were obtained with the {\tt StarTrack} population synthesis  code 
(Belczynski et al. 2002, 2008), with the inclusion of crucial updates in the physical 
models (winds, common envelope, supernovae). We have chosen several astrophysically
motivated models to illustrate the effects of some evolutionary processes on the 
formation of a NS-NS/BH-NS/BH-BH binaries. 

In particular, we explore the effects resulting from common envelope physics and 
uncertainties in compact object formation (including the supernova mechanism and natal 
kicks), as well as the effects of metallicity enrichment evolution through cosmic time. 
Our models use standard initial conditions for population synthesis calculations of 
massive binaries: thermal distribution of eccentricity ($\propto 2e$), moderate binary 
fraction ($50\%$), flat in $\log$ distribution of orbital separations ($\propto 1/a$), 
and uniform mass ratio distribution. Recently, Sana et al. (2012) have delivered a set 
of revised constraints on initial conditions for massive O stars: mostly circular 
binaries ($\propto e^{-0.42}$), high binary fraction ($>80\%$), preferentially short 
orbital periods ($\propto p^{-0.5}$), and uniform mass ratio distribution. However, 
Kobulnicky \& Fryer (2007) and later Kobulnicky et al. (2014) used limits from 
observations of massive stars and have not found similarly strong constraints; instead, 
they argue that the orbital parameters are still consistent with the old formulations.
Regardless, de Mink \& Belczynski (2015) demonstrated that the use of the Sana et al. 
(2012) initial conditions for massive binaries does not change (within a factor of $2$) 
the double compact object merger rates, and the resulting changes in NS and BH masses 
are negligible. 

Initial LIGO/Virgo have established upper limits on merger rate densities in the local 
Universe (Abadie et al. 2012; Aasi et al. 2013a). In what follows we make three 
successively refined estimates for the merger rate density sourced from population 
synthesis calculations to compare with these existing upper limits and with forthcoming 
upper limits from the advanced detectors.

\subsection{Method I}
\label{method1}

In the simplest approach one combines high- and low-metallicity models to evaluate 
Milky Way equivalent galaxy (MWEG) merger rates. Assuming a constant density of MWEGs,
the merger rate density may be estimated in the local Universe (Belczynski et al. 
2010a).

It is assumed that the local Universe has two components, with a fraction $f_{\rm Z}$ 
of local stellar density at solar metallicity ($Z=0.02$) with an associated 
Galactic merger rate ${\cal R}_{\rm MW}^{\zsunn}$, and the remaining $(1-f_{\rm Z})$ 
of stellar density at low metallicity ($Z=0.002$) with an associated merger rate 
${\cal R}_{\rm MW}^{0.1\zsunn}$. We do not use the Asplund et al. (2009) revision
of solar metallicity ($Z_\odot=0.014$). A recent comprehensive analysis of helioseismic 
and solar neutrino data indicates that this revision is not required (Villante et al. 
2014). We use the Galactic merger rates ${\cal R}_{\rm MW}^{\zsunn}$, 
${\cal R}_{\rm MW}^{0.1\zsunn}$, in units of [Myr$^{-1}$], from Dominik et al. (2012). 
We convert these into merger rate densities in the local Universe via
\begin{equation}
{\cal R}_{\rm vol} = 10^{-6} \rho_{\rm gal} \left( f_{\rm Z} 
{\cal R}_{\rm MW}^{\zsunn} + (1-f_{\rm Z}) {\cal R}_{\rm MW}^{0.1\zsunn} \right)
{\rm Mpc}^{-3} {\rm yr}^{-1}\, ,
\label{sfr1}
\end{equation} 
where we take the local density of Milky Way-like galaxies to be $\rho_{\rm gal}=0.0116$ 
Mpc$^{-3}$ (e.g., Kopparapu et al. 2008). In this calculation a constant star 
formation rate was assumed for MWEGs at a level of $3.5\msy$ for a duration of $10$ Gyr (this 
results in approximately the mass found in stars in the present day Milky Way). Only double 
compact objects that are formed with delay times (time elapsed from their formation
on the Zero Age Main Sequence to their merger) shorter than $10$ Gyr (the age of the
Galactic disk) contribute to the merger rate density.

The investigation of $\sim 30,000$ Sloan Digital Sky Survey galaxies revealed that 
recent (within the last $\sim 1$ Gyr) star formation was bimodal, with about half
of the stars formed with high metallicity, and the other half with low metallicity 
(Panter et al. 2008). We thus use a $50\%$--$50\%$ combination of two stellar 
populations, one with high and one with low metallicity ($f_{\rm Z}=0.5$).

Note that this method mostly ignores the star formation rate (SFR) evolution through 
cosmic time. It assumes a constant star formation rate per MWEG, so there is no merger
density variation as a function of distance/redshift, and it uses a very crude 
approximation of metallicity evolution over the history of the Universe, characterized by 
only two discrete metallicities. We also ignore selection effects in the GW detection 
of the binaries, such as accounting for how a given detector's reach depends upon 
binary total mass.

\subsection{Method II}
\label{method2}

The second method takes into account the star formation history and metallicity evolution 
through cosmic time. It also incorporates the full gravitational radiation waveforms and 
detector sensitivity noise curves. This method utilizes groundwork and models developed in 
Dominik et al.~(2013, 2015).  

The calculation begins with a SFR model across cosmic time (we adopt the extinction 
corrected model from Strolger et al. 2004) and metallicity evolution models. The 
metallicity evolution is only weakly constrained, especially for large redshifts, and 
therefore Dominik et al. (2013) have employed two models differing by the fraction of 
low-metallicity stars as a function of redshift. At each redshift there is a spread 
of metallicity of the stars formed in each model. The rate at which the average 
metallicity increases with cosmic time is constant in both models. At redshift $z=0$ 
the low-metallicity model results in a median metallicity of $0.8\zsun$, while for the
high-metallicity model the median is at $1.5\zsun$. In a nutshell, at any given 
time (or redshift) the average metallicity of star forming gas differs by a factor of 
$\sim 2$ between the two models. 

Population II and Population I stars are evolved with population synthesis over a large 
redshift range ($z=0$--$20$) and metallicity range ($Z=0.0001$--$0.03$). Double compact objects 
are formed and propagated in time to their merger point. As a result we obtain self-consistent 
merger rate densities as a function of redshift for NS-NS, BH-NS and BH-BH binaries 
(see Figs.~3 and~5 of Dominik et al. 2013 or \url{http://www.syntheticuniverse.org}). 
In what follows we present a method to evaluate which mergers are within the volume sampled 
by a GW detector of a given sensitivity. 

We assume that a given event is ``detected'' when it satisfies our detectability criterion: 
the source surpasses a signal-to-noise ratio (SNR) threshold in a single GW detector:
\begin{equation}
\mbox{SNR}>8.0.
\label{crit2}
\end{equation}
This single detector criterion approximately translates into $\mbox{SNR}>12$ for a 
3-detector network (e.g., two LIGO and one Virgo detectors). This is conventionally 
used as the threshold for a GW detector network to be able to identify the GW signal 
from a merging binary (e.g., Abadie et al. 2010). 

For the initial LIGO/Virgo observations we use a simple analytical approximation to the 
noise power spectral density given in eq.~3.1 of Ajith \& Bose (2009; see also Tab.~I 
of Sathyaprakash \& Schutz 2009).
For advanced LIGO/Virgo we adopt a noise model from an analytical approximation to 
the advanced LIGO zero-detuning high power noise power density of Ajith (2011; see their 
eq.~4.7). This approximation allows for effective observations above $\sim 20$ Hz and 
is in excellent agreement with the official advanced LIGO design noise curve 
(Shoemaker et al. 2010).  

Initial and advanced LIGO/Virgo rate estimates are obtained using gravitational waveform 
models for a given source with binary component masses $m_1$  and $m_2$ at redshift $z$. 
Our standard waveform model IMRPhenomC from Santamaria et al. (2010) includes inspiral, 
merger, and ringdown, and is tuned to numerical relativity simulations of non-precessing 
BH-BH mergers with aligned spins. We have verified explicitly that we obtain essentially 
identical rate estimates using an effective-one-body waveform model (more specifically, 
EOBNRv2; Pan et al. 2011). Mild differences between EOBNRv2 and IMRPhenomC only show up 
in the highest mass bins, where the rates are so low that statistical fluctuations dominate 
over uncertainties due to the gravitational waveform model. Pannarale et al. (2013) 
demonstrated that finite-size effects introduce negligible errors ($\lesssim 1\%$) in SNR 
calculations for BH-NS binaries, and therefore the IMRPhenomC model is also applicable to 
BH-NS rate estimates. We also use the IMRPhenomC waveforms for NS-NS binaries, since in 
this case the signal seen by advanced detectors is dominated by the early inspiral, where 
finite-size effects are negligible. The IMRPhenomC model includes higher harmonics in the 
waveform amplitude, resulting in a slight reduction of the inspiral amplitude compared to 
a simple quadrupole-formula estimate, and therefore in a slight decrease in the predicted 
detection rates. All of our rate calculations neglect spins, and therefore they should 
be considered as conservative lower limits (cf. Dominik et al. 2015 for a more detailed 
discussion of waveform models and of the effect of spins).

The detectability criterion of eq.~\ref{crit2} was applied to all population synthesis 
models in Dominik et al.~(2013). Mergers satisfying this criterion were selected and can 
be found online at \url{http://www.syntheticuniverse.org} following the link to ``Double
Compact Objects'' and then the link to ``Upper Limits''). In particular, these files 
include a list of merger events that are potentially detectable (i.e., are within the 
horizon redshift) of the initial and advanced LIGO/Virgo detectors. For each merger we 
list the intrinsic source-frame component masses ($m_1$ and $m_2$), the redshift of the 
merger ($z$), the SNR value for this binary assuming it is {\em optimally located and 
oriented}, the horizon redshift for this binary ($z_{\rm hor}$), and the merger rate 
density associated with a given event in the rest frame of the merger ($s$). This merger 
rate density is expressed in comoving volume units and refers to time as measured by a 
clock at the merger; it has units of Mpc$^{-3}$yr$^{-1}$. The horizon redshift indicates 
the redshift at which a particular binary merger with redshifted mass: $m_1(1+z_{\rm hor})$ 
and $m_2(1+z_{\rm hor})$ would have $\mbox{SNR}=8.0$ if it were optimally located and 
oriented with respect to a given detector. We then follow the method described below to 
calculate the merger rate density within the volume sampled by a GW detector of a given 
sensitivity.

We adopt a standard flat cosmology with 
$H_0=70.0\,\mbox{km}\,\mbox{s}^{-1}\,\mbox{Mpc}^{-1}$, $\Omega_m=0.3$, and $\Omega_\Lambda=0.7$ 
(and thus $\Omega_{\rm k}=0$) which is a good approximation to the latest estimates 
of cosmological parameters (Planck Collaboration 2015). We emphasize that none of our 
results depend sensitively on these parameters. 
The following relations, adopted from Hogg (2000), are used 
in our estimates. We present them explicitly for definiteness and to establish our notation. 
The relationship between redshift and (lookback) time is given by
\begin{equation}
t(z)=t_{\rm H} \int_{z}^{\infty} {dz' \over (1+z') E(z')},
\label{z2t}
\end{equation}
where $t_{\rm H}=1/H_0=13.969$ Gyr is the Hubble time, and
$E(z)=\sqrt{\Omega_{\rm M}(1+z)^3+\Omega_{\rm k}(1+z)^2+\Omega_\Lambda)}$.
The resulting age of the Universe is $t(0)=13.47$ Gyr. 
The comoving volume element $dV_{\rm c}/dz$ is given by 
\begin{equation}
{dV_{\rm c} \over dz}(z) = {c \over H_0} {D_{\rm c}{}^2 \over E(z)},  
\label{dVdz}
\end{equation}
where $c$ is the speed of light in vacuum, and where the comoving distance 
$D_{\rm c}$ is given by
\begin{equation}
D_{\rm c}(z) = {c \over H_0} \int_{0}^{z} {dz' \over E(z')}.
\label{Dc}
\end{equation}
From the comoving distance we can easily compute the luminosity distance ($D_{\rm l}$) for 
our adopted model of cosmology (with $\Omega_{\rm k}=0$):  
\begin{equation}
D_{\rm l}=(1+z) D_{\rm c}.
\label{Dl}
\end{equation}

Every merger from our population synthesis simulation is a proxy for a certain 
merger rate density:
\begin{equation}
s_i= \frac{d \mbox{SFR}}{d Z}\Delta Z\frac{1}{M_{sim}} \ 
{\rm Mpc}^{-3}\,{\rm yr}^{-1}, 
\end{equation}
where $\frac{d \mbox{SFR}}{d Z}\Delta Z $ is the fractional star formation rate in the 
simulated metallicity interval, $M_{\rm sim}$ is the total mass of single and binary 
stars (within our adopted initial mass function range: $0.08$--$150\msun$) in the 
simulation and star formation rate (SFR) is adopted from extinction
corrected model of Strolger et al. (2004).
 If a given merger from the population synthesis model satisfies eq.~\ref{crit2}, 
then the merger rate density which it represents, $s_{\rm i}$, contributes to the total 
merger rate density. The contribution to the merger rate (in the observer frame) associated 
with one particular simulated merger $i$ is
\begin{equation}
r_{\rm i} = 4 \pi p_{\rm det}\left({8.0 \over \mbox{SNR}}\right) s_{\rm i} 
{1 \over 1+z} {dV_{\rm c} \over dz} {dz \over dt}  \Delta t \ {\rm yr}^{-1},
\label{ri2}
\end{equation}
where the factor $1/(1+z)$ transfers the merger rate density $s_{\rm i}$ from the rest 
frame to the observer frame, the factor $dV_{\rm c}/dz$ is the comoving volume element,
the factor $(dz/dt) \Delta t$ allows us to integrate in time rather than redshift, and 
the factor of $4\pi$ takes into account the entire sky (i.e., integration over the solid 
angle). The population synthesis predictions were performed in finite time bins of 
$\Delta t=100\myr$. Both $dz/dt$ (eq.~\ref{z2t}) and $dV_{\rm c}/dz$ (eq.~\ref{dVdz}) are 
evaluated at the merger redshift. Finally, $p_{\rm det}(w)$ is a detection probability 
(with value in the range $0$--$1$) that takes into account the detector antenna pattern. 
This factor translates from $s_{\rm i}$ -- which represents the density within the entire 
spherical volume enclosed by the horizon redshift -- to the merger rate density within 
the ``peanut-shaped'' volume sampled by the detectors. To calculate this factor we use the 
approximation from the Appendix of Dominik et al. (2015):
\begin{multline}
p_{\rm det}(w)=
a_2 (1-w/\alpha)^2 + a_4 (1-w/\alpha)^4 + a_8 (1-w/\alpha)^8 + \\ 
(1-a_2-a_4-a_8) (1-w/\alpha)^{10}, 
\label{pdet}
\end{multline}
where $a_2=0.374222$, $a_4=2.04216$, $a_8=-2.63948$ and $\alpha=1.0$.
Each $r_{\rm i}$ is the contribution to the merger rate from a particular event $i$, and 
by summing over all such contributions our estimate takes into account the antenna pattern 
of the GW detectors. 

We calculate merger rate densities within a set of redshifted total
merger (binary) mass bins. 
Each bin includes some number of events ($n$), where each event is associated with its 
own specific merger mass and merger redshift. GW detectors do not provide measurements 
of the intrinsic merger masses, but instead are sensitive to the redshifted merger 
masses\footnote{It may be possible to infer a redshift distribution from the distance 
posteriors of a given event, and thereby infer a distribution over intrinsic merger mass. 
This is unlikely to be a productive exercise until significant numbers of detections 
arise, and statistical analyses can be performed.}. A merger is described (and assigned 
to a given mass bin) by its redshifted mass, as measured at the detector. The total 
redshifted mass of a binary merging at redshift $z$, as observed today ($z=0$), is 
defined by
\begin{equation}
M_{\rm tot,z}=(m_1+m_2)(1+z), 
\label{Mtotr}
\end{equation}
where $M_{\rm tot,i}=(m_1+m_2)$ is the intrinsic total mass of a given
merger. These cosmological factors will become increasingly important
as advanced detectors begin to probe the Universe to significant redshifts 
(up to $z \sim 2$ for heavy BH-BH mergers), implying that the most distant 
binaries are less massive than they appear to be in our detectors. 
Our results could easily be translated into chirp ($M_{\rm chirp}$) and 
redshifted chirp mass ($M_{\rm chirp,z}$):   
\begin{equation}
M_{\rm chirp,z}=M_{\rm chirp} (1+z) = {(m_1m_2)^{3/5} \over (m_1+m_2)^{1/5}} (1+z). 
\label{Mchirp}
\end{equation}

The total merger rate for a given redshifted mass bin within the volume 
sampled by a given detector is then calculated from 
\begin{equation}
r_{\rm tot} = \sum_{i=1}^{n} r_{\rm i} \ {\rm yr}^{-1}.  
\label{rtot}
\end{equation}
The $n$ events are generated, via population synthesis, from the entire cosmic 
star formation history (Population I and II; Population III is not included as
discussed in Dominik et al. 2013). Note that summing up merger rates from a 
range of redshifts, as we have done above, results in the loss of information 
about the redshift dependence of the merger rate. Note also that both $r_{\rm i}$ 
and $r_{\rm tot}$ are {\em detection} rates. 

The total merger rate density for a given redshifted total mass bin within the 
peanut-shaped comoving volume sampled by a given detector may be calculated from 
\begin{equation}
{\cal R}_{\rm tot} = \sum_{i=1}^{n} {r_{\rm i} \over V_{\rm antenna,i}} \ {\rm Mpc}^{-3} {\rm yr}^{-1},
\label{Rtot}
\end{equation}
where $V_{\rm antenna,i}$ indicates the comoving volume within a peanut-shaped
(antenna) volume for the specific merger event. For a given merger with component 
masses $m_1$ and $m_2$ and the corresponding horizon redshift $z_{\rm hor,i}$,  we 
estimate the comoving volume from: 
\begin{equation}
V_{\rm antenna,i}  = 4\pi \int_0^{z_{\rm hor,i}} {1 \over 1+z} {dV_{\rm c} \over dz}
p_{\rm det}\left({D_{\rm l}(z) \over D_{\rm l}(z_{\rm hor,i})}\right) dz \ {\rm Mpc}^{3},
\label{Vanti}
\end{equation}
where all factors have been defined earlier in this section.

Note that previous estimates (e.g., method I) neglect the fact that the merger rate 
density may change with redshift. In particular, for many models the merger rate density 
increases with increasing redshift (e.g., see Dominik et al. 2013; Figs.~3 and 5). 
This increase may be quite significant (factor of $\gtrsim 10$) within the volume 
sampled by advanced GW detectors (the most massive BH-BH mergers can be seen from as far 
as redshift of $z=2$; see Sec.~\ref{methods} and \ref{results2}). Therefore, the use of 
a constant merger rate density at any given redshift (e.g., $z=0$) is a major 
over-simplification. Estimates of the merger rate density must be integrated over the 
volume sampled by a specific detector for all sources. 

There are two crucial factors in the evaluation of merger rates or merger rate densities 
from modern population synthesis calculations that take into account SFR and metallicity 
evolution with redshift: a proper accounting of cosmology, and the inclusion of population 
synthesis results out to the horizon (and not just the range) of the detectors. We now 
elaborate on these two points in turn.

First, the effective volume included in the peanut-shaped antenna pattern sampled by a 
GW detector {\em cannot} be reliably calculated with simple formulas that utilize the 
Euclidean approximation. For example, initial LIGO/Virgo estimates typically
used the so-called range distance and volume:
\begin{equation}
D_{\rm range}= {D_{\rm hor} \over 2.264}, 
\label{Drange}
\end{equation}
\begin{equation}
V_{\rm range}={V_{\rm hor} \over (2.264)^3} = {V_{\rm hor} \over 11.605},  
\label{Vrange}
\end{equation}
where $\omega=2.264$ is the reduction factor that takes into account random sky positions 
and source orientations (i.e., ``sky and inclination averaged''; see, e.g., Eq.~(6) of 
O'Shaughnessy, Kalogera, \& Belczynski 2010), $D_{\rm hor}$ is the 
luminosity distance corresponding to the horizon redshift, and $V_{\rm hor}$ is the entire 
(spherical) volume enclosed within the horizon redshift. 
For example, the volume within the horizon redshift $z_{\rm hor}=2$ (relevant for 
advanced detectors; see Sec.~\ref{results2}) is $V_{\rm hor}=58.0 \times 10^{10}$ Mpc$^3$ 
(integral of eq.~\ref{dVdz}). 
This leads to a range volume of $V_{\rm range}=5.0 \times 10^{10}$ Mpc$^3$ using
eq.~\ref{Vrange}, and to a much larger value of $V_{\rm range}=16.2 \times 10^{10}$ Mpc$^3$ 
using eq.~\ref{Drange} (where $D_{\rm range}$ is used to calculate the range redshift 
$z_{\rm range}$ via eq.~\ref{Dl}, and this in turn is used to calculate the range volume 
from the integral of eq.~\ref{dVdz}). Taking into account cosmology, the correct expression 
is eq.~\ref{Vanti}, which produces a volume of $V_{\rm antenna}=7.1 \times 10^{10}$ Mpc$^3$.

Second, when using population synthesis results which produce merger rates that vary with 
redshift, it is necessary to account for all sources (alas with diminishing probability 
defined by $p_{\rm det}$) all the way out to the horizon redshift; the inclusion of events 
only out to the range redshift is insufficient. Redshift evolution of merger rates is fully 
expected: these rates are predicted to change by a factor of $\sim 10$ in the redshift 
range $z=0$--$2$ (Dominik et al. 2013).

In what follows we abstain from the use of the term ``range'' in our calculations and text, 
and the term ``antenna'' to denote the volume sampled by a given (initial or advanced) 
detector.

Our method utilizes the astrophysical knowledge of the sources (single and binary 
evolution in population synthesis models) combined with the cosmological evolution 
of the SFR and metallicity, while also incorporating the full waveforms 
(inspiral-merger-ringdown) from the merging sources and the detailed detector noise 
curves. There are of course many uncertainties, some very important, in the underlying 
evolutionary models. Furthermore, there are minor uncertainties introduced by unknowns 
associated with the cosmological parameters, the waveforms, and the future noise curves 
of the advanced LIGO  detectors. Although specific parts of our analysis may be updated 
in the future with improved and revised input physics, our basic framework for calculating 
the merger rate density within a cosmological context should remain applicable.

There is one potential issue that should be taken into account while working with GW 
observations (whether these are upper limits or detections). Observational imperfections
or statistical GW measurement errors (due to noise fluctuations) will affect any comparison 
with theoretical models. This issue may be addressed in two different ways: corrections may 
be applied to the population synthesis model predictions, or the GW detection errors may be 
incorporated into observations and presented as uncertainties on the measured upper 
limits/detection rates. The practical application of the former approach is presented in 
Stevenson, Ohme, \& Fairhurst (2015), while here we adopt the latter approach and assume 
that the observational uncertainties are appropriately accounted for.

\subsection{Method III}
\label{method3}

In method III we compute merger rate densities based on the {\em entire} spherical 
volume enclosed by the horizon redshift, and not just within the peanut-shaped detector 
antenna pattern as was pursued in method II. In the following we only rewrite those 
equations which are modified with respect to method II (see Sec.~\ref{method2}).

The observer frame merger rate associated with one particular event $i$ is:
\begin{equation}
r_{\rm i} = 4 \pi s_{\rm i} {1 \over 1+z} {dV_{\rm c} \over dz} 
{dz \over dt}  \Delta t \ {\rm yr}^{-1}, 
\label{ri3}
\end{equation}
where all the factors have already been introduced in Sec.~\ref{method2}. This is the 
contribution to the merger rate from a particular event $i$, and we will sum all these 
contributions all the way to the horizon distance for each observed mass bin. Note that 
the detection probability factor $p_{\rm det}$ does not appear in eq.~\ref{ri3}.

The total merger rate density for a given redshifted total mass bin within spherical 
comoving volume limited by the horizon redshift is given by 
\begin{equation}
{\cal R}_{\rm tot} = {r_{\rm tot} \over V_{\rm hor}} \ {\rm Mpc}^{-3} {\rm yr}^{-1},
\label{Rtot3}
\end{equation}
where $V_{\rm hor}$ indicates the comoving volume within the horizon redshift 
$z_{\rm hor}$ for a given redshifted total mass bin. This quantity is estimated by 
integrating eq.~\ref{dVdz}:
\begin{equation}
V_{\rm hor}  = {4\pi \over 3} D^3_{\rm c}(z_{\rm hor}),  
\label{Vhor}
\end{equation}
where $D_{\rm c}$ is given by eq.~\ref{Dc}.

If the merger rate density were constant (from the point of view of the observer frame, 
so constant in redshifted mass and redshifted time bins) out to the horizon redshift, 
$z_{\rm hor}$, then the inferred rate densities within the peanut (method II) and within 
the full volume (method III) would be identical.
However, it would be a very striking coincidence if the merger rate density of double 
compact objects is constant out well beyond the Hubble flow. Double compact object 
formation depends on the star formation rate and on various properties of the stars that 
form compact objects (e.g., metallicity, binarity, IMF). It is well established that at 
least some of these quantities/properties change with redshift (e.g., SFR and metallicity), 
and therefore it seems highly unlikely that these redshift dependent factors result in a 
constant merger rate density in the observer frame. The relevant question for our study is: 
how significantly does the merger rate density change when going from $z=0$ to the horizon 
redshift? 

A first intuitive estimate can be easily arrived at: if coalescence times for double compact 
objects are short, then the merger rate density will evolve with redshift in a similar fashion 
to the SFR. Within the horizon redshift for the most massive BH-BH mergers ($z_{\rm hor}=2$), 
the SFR increases by a factor of $\sim 10$, implying a similarly dramatic evolution in merger
rate. This intuitive expectation (modified somewhat by other factors, such as metallicity 
evolution with redshift) is confirmed by detailed population synthesis estimates (Dominik et 
al. 2013).  

LIGO/Virgo will sample merger rate densities (although with decreasing sensitivity) all the 
way out to the horizon redshift. In principle, knowing the antenna pattern and assuming a 
prior on the merger rate density redshift dependence, it is possible to infer from the number 
of observed binaries an estimate of the upper limit or merger rate density in the entire 
spherical volume enclosed within the horizon redshift. Moreover, in the case when there are 
plentiful detections, the rate can be estimated observationally for specific distance (or 
redshift) intervals.

\subsection{Comparison of Methods}
\label{methods}

In what follows we present the merger rate density divided into bins of redshifted total 
merger mass. Since GW detectors will measure the redshifted mass of mergers, we therefore use 
the redshifted total merger mass from our simulations (eq.~\ref{Mtotr}). The specific limits 
on the bins are adopted from the initial LIGO/Virgo upper limits papers for low- and high-mass 
inspirals (Abadie et al. 2012; Aasi et al. 2013a), with a highest mass bin consisting of 
$91$--$109\msun$. We extend this binning, with bin width of $18\msun$, out to a maximum of 
$500\msun$. This choice is arbitrary, and can be changed if desired.  

\begin{figure}
\vspace*{-3cm}
\includegraphics[width=1.0\columnwidth]{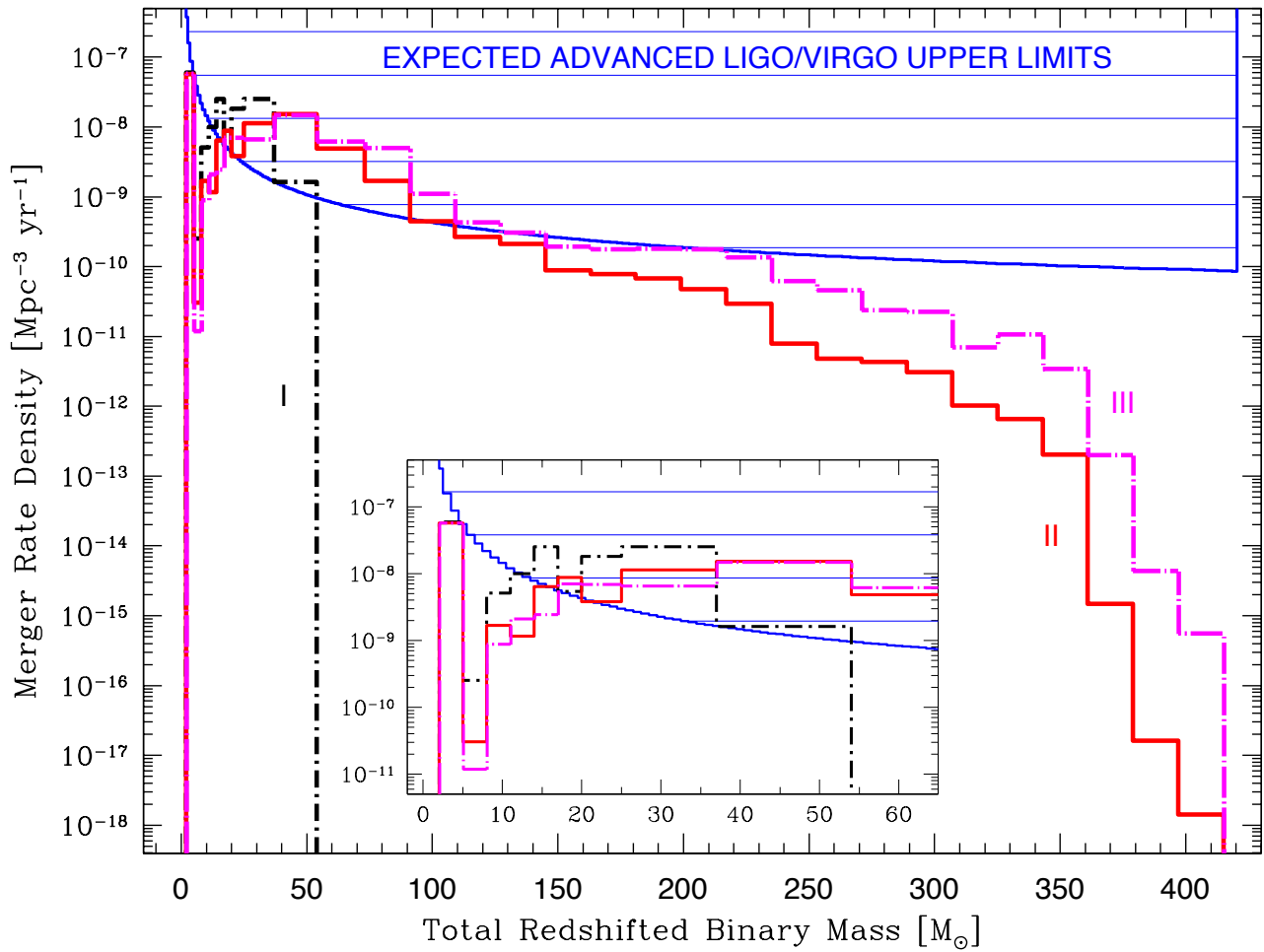}
\vspace*{-5.3cm}
\caption{
Comparison of methods for estimating the double compact object merger rate
density with advanced LIGO/Virgo. The rate densities are plotted for a range 
of total redshifted merger mass bins.
The theoretical predictions are shown for one underlying population 
synthesis evolutionary model (V2l). The three lines demonstrate the dependence
of the rate density on the calculational method:
method I (black short dashed-dotted line), method II (red solid line), 
and method III (magenta long dashed-dotted line).
Method II is the correct prediction for the merger rate densities as measured by
advanced LIGO/Virgo. Method I ignores cosmology, and therefore fails
dramatically at higher masses and distances.
Variants of this method have been  generally used 
within the population synthesis community to generate rate predictions for double compact object 
merger rates; for example, all the detection probabilities/merger rates presented within 
Abadie et al. (2010) use Method I or even simpler approaches.
Method III ignores the antenna power pattern of the detectors, and thereby
utilizing more volume to estimate rates and overestimates the LIGO/Virgo rates
at high mass.
The blue line shows the expected advanced LIGO/Virgo upper limits. The inset zooms in 
on mergers with relatively low total mass. 
Places where the predictions overlap significantly with the upper
limit curve indicate that the models predict detections.
The redshifted total mass of potentially detectable binaries can reach $400\msun$ 
(corresponding to a redshift $z=2$ merger with a total intrinsic mass of $130\msun$). 
}
\label{f1}
\end{figure}

In Figure~\ref{f1} and Table~\ref{t1} we present the dependence of the merger rate
density on the calculational method (method I, II, and III). For this exercise we
have used one specific population synthesis model, so as to highlight the differences 
in our merger rate density calculations. In methods II and III we employ the evolutionary 
model V2 with low metallicity enrichment; this model utilizes restricted CE survival 
(see Sec.~\ref{models}). In method I we use the same evolutionary model V2, but the 
metallicity evolution treatment is oversimplified. Additionally, in this case no merger 
redshift information is available, so only the intrinsic total merger masses are used. 
Obviously, if we assume---as we have---a constant comoving density of Milky Way-like 
galaxies (see eq.~\ref{sfr1}), we can draw from a random distribution of redshifts. However 
this modification is not applied to our calculations, as our intent is to emphasize the
differences between the methods that use (or do not use) redshift information about the 
mergers. We find strikingly large differences between method I and the other two 
methods of merger rate density calculation. 
For comparison we also include the expected advanced LIGO/Virgo upper limits (see 
Sec.~\ref{aligo}). We note that this curve represents the upper limits 
resulting from non-detection. When comparing these limits to the predicted 
rates, there are two relevant points of comparison: the overall rate, and the mass 
distribution. Places where the predictions overlap significantly with the upper limit 
curve indicate that the models predict detections. In these cases, the predicted event 
rate of BH-BH mergers across all mass bins would be inconsistent with an absence of  
detections in advanced LIGO (as represented by the upper limit curve). In addition, by 
comparing the shape of the upper limit curve to the shape of the predictions, we are 
provided with a visual indication of how the mass selection effects of the instruments 
impacts the predicted mass distributions of sources. 

Methods II and III generate very similar merger rate densities for relatively low mass 
mergers ($M_{\rm tot,z}<70\msun$). For higher total redshifted merger masses, method III 
predicts consistently higher merger rate densities than method II. This is expected, as 
by construction method III samples a larger high-redshift volume than method II. The 
BH-BH merger rate density increases with redshift within the entire advanced LIGO/Virgo 
horizon of about $z_{\rm hor}=2$ (see Sec.~\ref{methods} and ~\ref{results2}).

Methods II and III predict nonzero rates out to a total redshifted mass of $400\msun$.
This is a much higher total BH-BH binary mass than what is conventionally expected from 
normal stars, where by ``normal'' stars we mean stars with initial mass below $150\msun$. 
Until recently, this value was believed to be the upper mass limit on star formation, at 
least in non-zero metallicity environments. Thus far all the predictions for compact object 
merger rates from Population I and Population II stars have been limited to stars of mass 
$<150\msun$. The recent discovery of stars estimated to be initially as massive as 
$\sim 200$--$300\msun$ (e.g., Crowther et al. 2010) in the relatively high metallicity 
environment of the S Doradus cluster in the LMC ($Z \approx 0.6 \zsun$) has shaken these 
beliefs. The first study of BH-BH merger rates from these very massive stars has already 
been proposed (Belczynski et al. 2014). However, in the current work our models are limited 
to stars with initial mass below $150\msun$ and we nonetheless find mergers as massive as 
$400\msun$ in our calculations. Two factors are responsible for this surprising result: the 
inclusion of very low-metallicity models, and our use of the full gravitational waveform in 
our analysis. The former allows for the formation of massive BH-BH binaries from Population 
II stars even if these are limited to $150\msun$, while the latter allows for the detection 
of high mass BH-BH mergers from large distances, thereby redshifting the intrinsic total 
merger mass to higher values (see eq.~\ref{Mtotr}). 

For the evolutionary models we use in this study, the most massive BH-BH binary is formed at 
our lowest adopted  metallicity ($Z=0.0001$) from two stars with very high initial (Zero Age 
Main Sequence) mass: $m_{\rm zams,1}=148\msun$ and $m_{\rm zams,2}=144\msun$. The detailed 
evolutionary sequence is described in Dominik et al. (2013; see their Sec.~5.1) and it ends 
in the formation of a BH-BH binary with a total intrinsic BH-BH mass of 
$M_{\rm tot,i}=136\msun$ (component masses of $74\msun$ and $62\msun$). This is not surprising 
within the adopted evolutionary framework, as wind mass loss is ineffective at such low 
metallicity {\em  and} the most massive BHs form via direct collapse without a supernova event
(and its associated mass loss). This highlights the physically motivated possibility of
such high mass scenarios, given our assumptions. Our application of the full waveform 
for such a merger allows us to state that the merger, with a redshifted mass of 
$M_{\rm tot,z} \simeq 408\msun$, remains detectable with the advanced LIGO detectors out 
to a redshift of $z=2$. Such high redshifts are not generally considered accessible in the 
context of GW observations with second generation instruments, but if these sources exist 
they may be potentially detected at these vast distances. Our evolutionary predictions 
place such high mass ($\sim 400 \msun$) events at very low merger rate density; $\sim 5$--$8$ 
orders of magnitude below forecasted advanced LIGO/Virgo upper limits. However, at 
somewhat smaller total redshifted merger mass ($\sim 300$--$350\msun$) the predictions
are only $\sim 1$-$3$ orders of magnitude below upper limits making detections feasible. This 
is especially true if we consider $10$ years of advanced LIGO/Virgo observations. Note also 
that such massive mergers could result from other formation mechanisms (Belczynski et al. 
2014).

The merger rate densities predicted with method I are very different from those obtained 
from the more physical and self-consistent approaches in methods II and III. The only 
agreement is found for NS-NS binaries, for which all three methods agree. However, for BH-NS 
and BH-BH binaries there is striking disagreement: for low mass binaries ($5$--$35\msun$) 
method I produces a merger rate density that is $\sim 5$--$10$ times higher than from 
method II or III, and for higher mass binaries the method I merger rate density quickly 
drops below the rates from method II and III. 

Method I does not produce any mergers with total intrinsic mass above $50\msun$, while 
methods II and III allow for non-zero merger rate densities of systems with total observed 
mass up to $400\msun$. Note that these striking differences are obtained with the same 
underlying evolutionary model and therefore can be fully associated to {\em (1)} the 
differences in calculation methods, {\em (2)} the much broader range of metallicity in 
methods II and III ($Z=0.0001$--$0.03$) as compared to method I ($Z=0.002$--$0.02$), and 
{\em (3)} the fact that there is no redshift information available in method I, so the merger 
mass is not redshifted. However, even if we apply a redshift to the total mass of {\em all} 
the mergers in method I using {\em the most extreme} redshift of potentially detectable 
binaries  ($z=2$), the maximum redshifted mass in our suite of simulations from method I 
would still only reach $150\msun$. 

The remaining part of the difference comes from the fact that method I uses only two 
metallicities ($Z=0.002$ and $Z=0.02$) while methods II and III incorporate a much 
broader spectrum of metallicity ($11$ values spanning the range $Z=0.0001$--$0.03$). 
Metallicity is a very important factor in the formation of BH-BH mergers (Belczynski 
et al. 2010a). In a nutshell, decreasing metallicity decreases wind mass loss and can 
increase the chance of common envelope development/survival, thereby enhancing the 
formation of close and massive BH-BH binaries. As a result, the maximum intrinsic total 
mass of a BH-BH binary is $M_{\rm tot,i}=42\msun$ ($24+18\msun$ system formed at 
metallicity $Z=0.002$) using method I, and $M_{\rm tot,i}=136\msun$ ($74+62\msun$ 
system formed at $Z=0.0001$) using methods II or III. The specific double compact object 
formation channels and the mass dependence of double compact objects on metallicity is 
discussed by Dominik et al. (2012; 2013).     

The above discussion argues against using simple methods, such as method I, to make 
comparisons with advanced GW observations. Note that merger and detection rate estimates 
for double compact objects (many of which come from past studies with the {\tt StarTrack} 
population synthesis code; e.g., Belczynski et al. 2002) collected in Abadie et al. (2010) 
use only a {\em single} value of metallicity ($Z=0.02$). It is important to note that this 
value strongly disfavors the formation of BH-BH mergers. Since metallicity can 
significantly alter the predicted rates, and since merger rate estimates must be framed in 
a cosmological context in the advanced detector era, the Abadie et al. (2010) estimates 
for the stellar-origin double compact object merger rates must be significantly updated.

In summary, we find that method I is significantly flawed, especially at higher mass, and 
we do not use it in what follows. We will use method II for {\em all} of our comparisons 
with the initial and advanced LIGO/Virgo upper limits.

\section{Evolutionary Models}
\label{models}

Following Dominik et al. (2013; 2015) we consider four evolutionary models
(see Table~\ref{t2} for model summaries).
The models employ the current best estimates for various physical parameters, 
including some that are not yet fully constrained but play an important role in the 
formation of double compact objects. For example, during CE evolution physical 
estimates of the donor envelope binding energy are used (Xu \& Li 2010; as revised 
by Dominik et al. 2012), we adopt $M_{\rm NS,max}=2.5 \msun$ as the maximum NS mass 
(Lattimer \& Prakash 2011), we assume the NS natal kick distribution, based
on observations to be a Maxwellian with $\sigma=265$ km s$^{-1}$ (Hobbs et al. 2005), 
BH natal kicks are smaller and due to a mass ejection mechanism (Fryer et al. 2012), 
the compact object mass spectrum is based on rapid supernova explosions (Belczynski 
et al. 2012), the stellar winds are revised for the effects of clumping (Vink, de 
Koter \& Lamers 2001) and constrained by black hole mass estimates (Belczynski et al. 
2010b), and we adopt non-conservative mass transfer episodes with $50\%$ of the mass 
retained in the binaries (Meurs \& van den Heuvel 1989). 

A detailed discussion of evolutionary channels and accompanying uncertainties is 
provided in Dominik et al.~(2012). For example, BH-BH binaries are formed 
predominantly~($>90\%$) along one very specific evolutionary channel: Two massive 
stars $20$--$150\msun$ begin evolution on a wide orbit ($\sim 1000 \rsun$).
The first binary interaction is a stable Roche lobe overflow (RLOF) that does not 
significantly change the orbit. The exposed core of the primary (initially more massive) 
star collapses directly to form a first BH. The second binary interaction is
initiated by the secondary, and the RLOF proceeds on the dynamical scale leading to 
CE evolution. In the process the orbital separation decays to $\sim 10 \rsun$. The 
exposed core of the secondary star then collapses to form a second BH. This BH 
formation also occurs without any supernova explosion and the system remains bound. 
A close BH-BH binary is formed with a coalescence time shorter than the Hubble time.
The two major uncertainties involve CE evolution and BH formation. 

In models V1 and V2 we test the potential suppression of the NS-NS/BH-NS/BH-BH formation 
in case B RLOF. If the RLOF proceeds on the dynamical timescale of 
the donor star a given binary evolves through a common envelope (CE) phase. The severe 
orbital decay during CE is one of the mandatory conditions in the formation of close 
double compact objects. However, it is not at all clear whether massive Hertzsprung gap 
(HG) donors (case B) have clear core-envelope structure (Belczynski et al. 2007) and 
whether they behave like MS stars (always resulting in a CE merger) or evolved giant 
stars (with potential CE survival). Additionally, the conditions for CE development are not 
fully understood (e.g., Ivanova et al. 2013). This is true in particular for massive stars 
that may have radiative envelopes during most of their HG life and may potentially evolve 
through thermal timescale RLOF rather than CE. Such evolution would not provide any 
significant binary orbit contraction, and it would result in wide NS-NS/BH-NS/BH-BH formation. 
Conditions for CE development are critical in the formation of close (merging) double 
compact objects, and are currently under study with the detailed evolutionary code MESA 
(Pavlovski, Belczynski, \& Ivanova, in preparation). In model V1 we adopt an optimistic 
scenario and allow HG stars to survive through the CE phase. In model V2 we remove from 
our simulations all potential NS-NS/BH-NS/BH-BH progenitors that encounter CE with an HG 
donor star; this is probably more physically realistic than model V1. We use model V2 as 
our standard (reference) model, and we use its relevant CE physics in subsequent models.

In models V3 and V4 we test supernova/core-collapse physics. In particular in model V3 
we adopt high BH natal kicks (drawn from a 1-D Maxwellian with $\sigma=265$ km s$^{-1}$).
Such high kicks are most often adopted for NSs. In all other models the BH natal kicks are 
much smaller and we describe these particular choices in more detail in Sec.~\ref{compkicks}. 
In model V4 we change the underlying supernova/core-collapse model, and the resulting 
NS/BH mass spectrum changes from one with a mass gap (models V1,V2 and V3) to a
continuous spectrum without a gap.  The mass gap (lack of compact objects in mass range 
$2$--$5\msun$) has been a puzzling observation for over a decade (Bailyn 1998). At present 
the origin of the gap remains unclear. It has been proposed that the gap is just an 
observational artifact (Kreidberg et al. 2012). Alternatively, the gap may be set by the 
very short time in the development of a supernova explosion in the neutrino supported 
convective engine model (Belczynski et al. 2012). 

For all models we use a typical Initial Mass Function (IMF) with slope of $-2.7$ for 
massive stars, $50\%$ binary fraction, thermal distribution for initial eccentricity,
and flat in logarithm initial distribution of initial separations (as described in 
detail in Dominik et al. 2012). We have assumed that the IMF ends at $150\msun$. As 
discussed above, this may no longer be true given the massive stars discovered in the 
R136 cluster in the LMC (Crowther et al. 2010). We have dedicated a separate study to 
investigate the formation of BH-BH binaries from such very massive stars (Belczynski 
et al. 2014). The initial distributions may also appear outdated; new measurements of
O-type stars indicated quite different distributions for orbital periods and 
eccentricities (Sana et al. 2012). However, adopting the new distributions does not
significantly affect the merger rates nor the masses of double compact object (de Mink 
\& Belczynski 2015).

\section{Observations}
\label{obsligo}

\subsection{Initial LIGO/Virgo upper limits}
\label{iligo}

The LIGO/Virgo S6 run (completed in 2010) combined with earlier searches (initial LIGO/Virgo) 
was used to derive upper limits on double compact object merger rate densities. For 
mergers with total mass below $25\msun$ the upper limits were provided in seven mass 
bins: $2$--$5\msun$, $5$--$8\msun$, $8$--$11\msun$, $11$--$14\msun$, $14$--$17\msun$, 
$17$--$20\msun$, $20$--$25\msun$ (Abadie et al. 2012; see their Fig.~4). This search 
was performed utilizing only the inspiral part of the waveform templates adequate for 
low mass mergers.  

For mergers with higher total mass the upper limits were provided as a function of the 
two component masses (Aasi et al. 2013a; see their Fig.~5 and Table 1). These upper limits
were obtained with full waveforms including inspiral, merger and ringdown. The deepest
upper limits correspond to equal mass binaries. For example, the upper limit for a 
detection of a $50$--$50\msun$ BH-BH merger is $0.7\times 10^{-11}\mpy$, which is a factor 
of $\sim 5$ lower than the $3.8\times 10^{-11}\mpy$ limit for the $77$--$23\msun$ merger. 
This is the most extreme case of upper limit difference for the same total mass but 
different mass ratio presented in Aasi et al. (2013a). Typically, the change from equal 
mass binary to uneven mass binary results in increasing upper limits by less than a factor 
of $2$. For example, the upper limit changes from $3.3\times 10^{-11}\mpy$ for a
$23$--$23\msun$ merger to $4.2\times 10^{-11}\mpy$ for a $32$--$14\msun$ merger (change of
only $\sim 1.3$).

The adoption of equal mass upper limits provides the most optimistic estimates and the 
closest approach with evolutionary model predictions for the merger rate density. Keeping 
this in mind, we note that our recent evolutionary predictions typically generate BH-BH 
binaries with comparable mass components (average mass ratio of $0.8$; see Fig.~9 of 
Dominik et al. 2012). On the other hand, BH-NS binaries are predicted to form with 
extreme mass ratios (average mass ratio of $0.2$; also see Fig.~9 of Dominik et al.
2012). However, it is not expected that their total mass would typically exceed 
$25\msun$. If this is the case, then most BH-NS mergers would contribute in bins 
covered by the low mass search, in which a uniform distribution of primary mass was adopted 
(Abadie et al. 2012). This makes these upper limits applicable to BH-NS mergers.

Technically, the borders of high mass bins adopted in our study were obtained from 
mid points between five equal mass binaries presented by Aasi et al. (2013a) along with 
the quoted S5+S6-Vsr2/3 EOBNR upper limits. 
For example, the first two equal mass binaries in Table~1 of Aasi et al. (2013a) are 
$14$--$14\msun$ ($M_{\rm tot}=28\msun$) and $23$--$23\msun$ ($M_{\rm tot}=46\msun$) with 
the mid point in total mass of $37\msun$. That gives us the first high mass bin: $25\msun$ 
(the end point of low mass search binning) to $37\msun$ (the mid point between the first 
two equal mass binaries in high mass search).

\subsection{Advanced LIGO/Virgo upper limits}
\label{aligo}

We have calculated the expected upper limits for the design advanced LIGO/Virgo 
sensitivity. As discussed near Eq.~\ref{crit2}, we conservatively assume that detection 
requires two-instrument coincidence. 
Our estimated sensitivity is therefore independent of the number of detectors in 
the network (so long as there are at least two) as well as their relative
locations and orientations. This assumption implies that 
{\em (a)} the network's sensitivity is set by the second-most-sensitive detector, 
here assumed to be one of the advanced LIGO/Virgo detectors operating in the zero 
detuned/high-power configuration (Aasi et al. 2013b); and {\em (b)} that for a calendar time 
$T\, {\rm yr}$ of operating time, the network accumulates only $p^2 T\, {\rm yr}$ coincident 
time, where $p=0.8$ is the duty cycle applied for the two most sensitive instruments. Using 
these assumptions, we anticipate that advanced LIGO/Virgo will quote upper limits as a function of 
\emph{detected} (redshifted) mass $M_{\rm tot,z}$ (in $M_\odot$) computed by the following 
expression: 
\begin{eqnarray}
{\cal R}_{aLIGO/Virgo} = \frac{2.3}{V p^2 T} {\rm Mpc^{-3} yr^{-1}}, 
\end{eqnarray}
where $V$ is the volume inside which a binary with the redshifted mass $M_{\rm tot,z}$ could 
be seen by the second-most sensitive detector, using the detection threshold $\mbox{SNR}=8$ 
(see Eq. \ref{crit2}). 
Accounting for cosmology and orientation-dependent sensitivity, we evaluate this volume for  
each $M_{\rm tot,z}$ by first computing the maximum luminosity distance 
$D_{\rm hor}(M_{\rm tot,z})$ to which the instrument is sensitive at that mass, using our 
detector sensitivity and a model for binary coalescence; using cosmology to find the 
associated horizon redshift $z_{\rm hor}$; and then evaluating the volume $V$ with 
eq.~\ref{Vanti}. 
To eliminate the ambiguity with spin and mass ratio, for simplicity we assume all coalescing 
binaries have equal mass and zero spin: we evaluate $D_{\rm hor}(M_{\rm tot,z})$ using an 
effective-one-body waveform model (EOBNRv2; Pan et al. 2011). This assumption is a very 
good approximation for BH-BH and NS-NS mergers (typical mass ratios of $q \sim 1$; Dominik et 
al. 2012), however it is not ideal for BH-NS mergers (typical mass ratios of $q \sim 0.2$). 

The upper limits obtained with the above detailed calculation can be contrasted with a simple 
intuitive estimate (see Appendix~\ref{detUL}).

\section{Upper Limits versus Predictions}
\label{results}

\subsection{Comparison of Models with Initial LIGO/Virgo Upper Limits}
\label{results1}

\begin{figure}
\includegraphics[width=1.0\columnwidth]{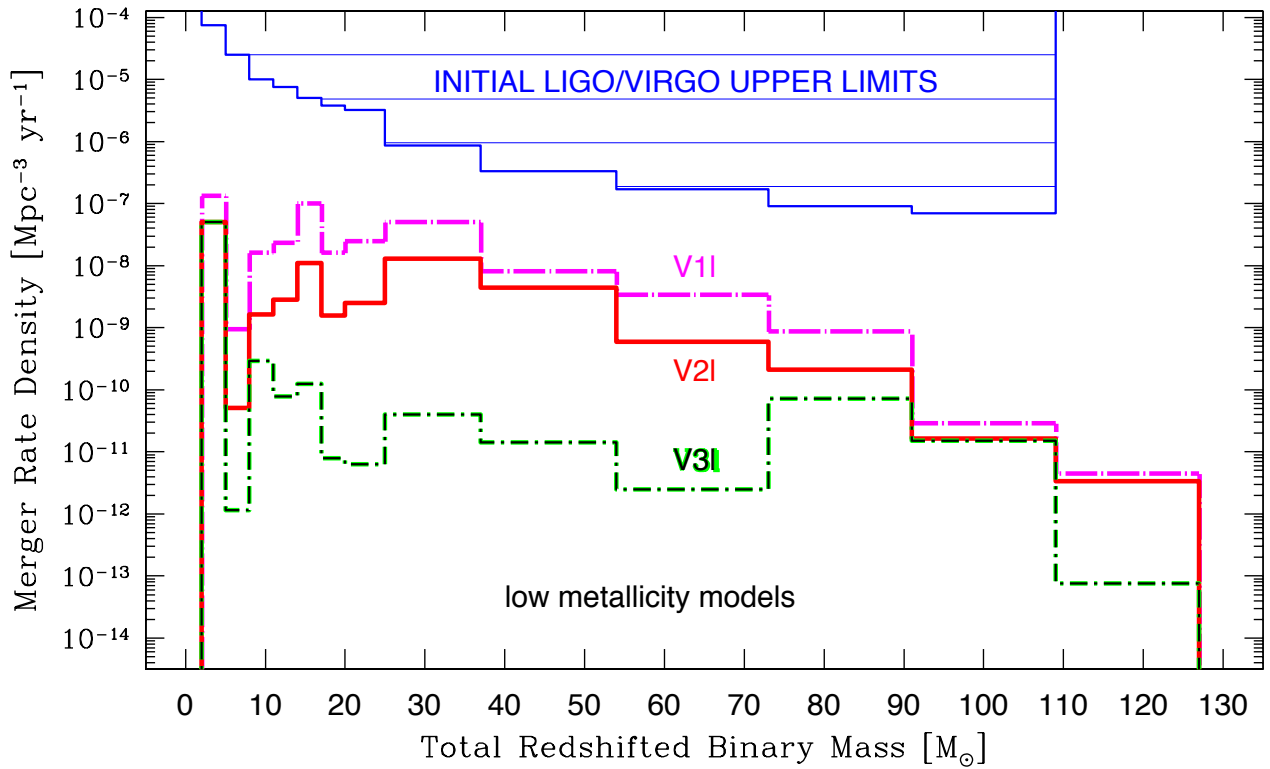}
\vspace*{-5.3cm}
\caption{
Merger rate density of double compact objects for our low metallicity
evolution scenario predicted (with method II) for initial LIGO/Virgo. Note 
that the predictions are very close to the upper limits (within a factor of $18$) for 
model V1 for total merger masses of $25$--$37\msun$. The merger rate density decreases 
from model V1 (optimistic CE), to V2 (standard binary evolution), to V3 (high BH 
natal kicks). 
Model V4 (not shown) results in a very similar merger rate density level to model 
V2, with the exception that the second mass bin ($5$--$8\msun$) does not
show a characteristic drop (mass gap) found in other models. The top shaded 
area shows the upper limits from initial LIGO/Virgo. 
}
\label{f2}
\end{figure}

\begin{figure}
\includegraphics[width=1.0\columnwidth]{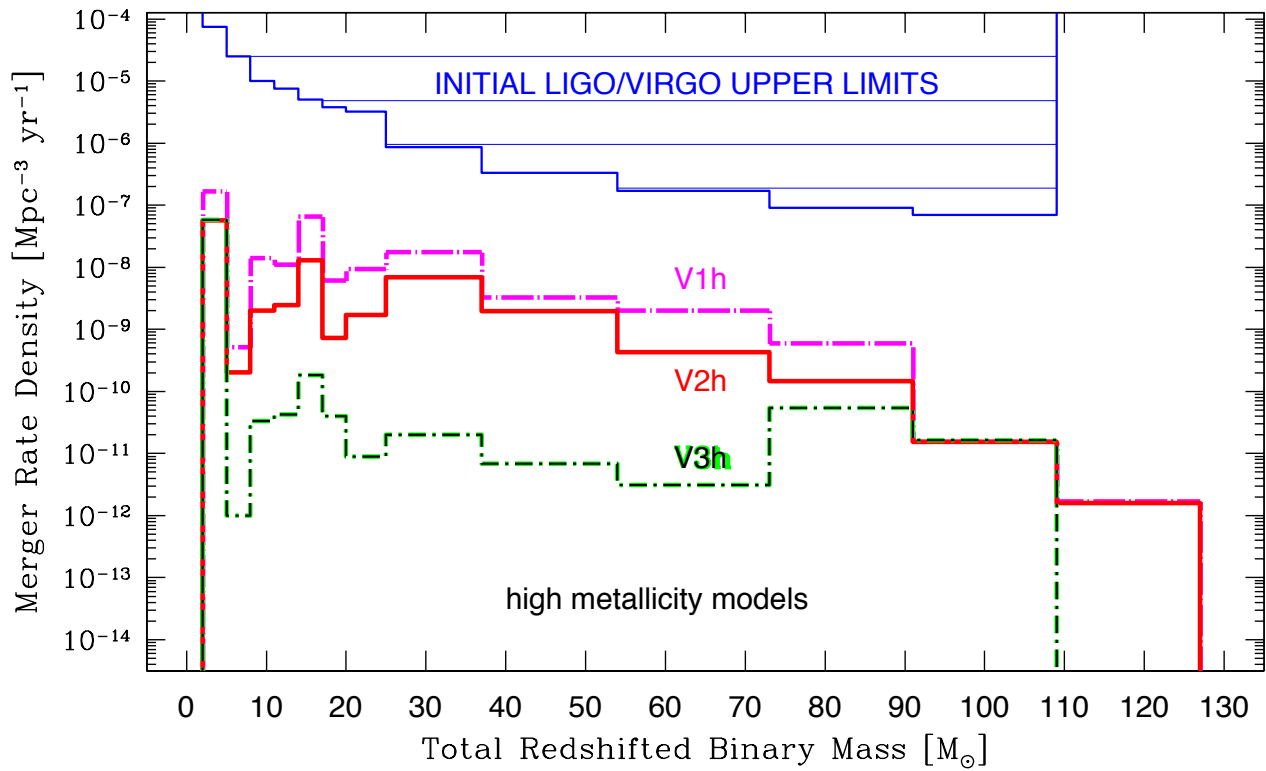}
\vspace*{-5.3cm}
\caption{
Merger rate density of double compact objects for our high metallicity
evolution scenario predicted (with method II) for initial LIGO/Virgo. 
Otherwise the same as Fig.~\ref{f2}.
Note that although the metallicity is a key factor in the formation of 
double compact object binaries, the variation of the cosmological metallicity 
evolution scenario does not significantly affect merger rate densities. 
For example, in model V2 the merger rate density decreases by less than a factor 
of $\sim 4$ in all mass bins when going from low to high metallicity evolutionary 
scenarios (see also Table~\ref{t2} and Table~\ref{t3}).  
}
\label{f3}
\end{figure}

In Figures~\ref{f2} and~\ref{f3} and Tables~\ref{t3} and~\ref{t4} we present evolutionary 
model predictions for double compact object merger rate densities. The predictions are 
constructed with method II (see Sec.~\ref{method2}) and contrasted with observational 
initial LIGO/Virgo upper limits (see Sec.~\ref{iligo}). Our models are described in
Sec.~\ref{models}. In what follows we list the most noteworthy trends, and compare directly 
predicted rate densities with the existing observational upper limits.

The dependence of the merger rate density on the total mass of the binary begins with 
a pronounced peak in the first mass bin ($M_{\rm tot,z}=2$--$5\msun$) where all of the 
NS-NS systems are found. It is possible that some BH-NS systems may contribute to this 
bin in model V4 (no mass gap). However, since our standard approach naturally 
produces a mass gap between NSs and BHs (BHs form with mass above $\sim 5\msun$; see 
Belczynski et al. 2012) there are no BH-NS systems in this first bin for models V1, V2, V3.  
The next mass bin ($M_{\rm tot,z}=5$--$8\msun$) with relatively low merger rate densities 
and the two following bins (with total mass $M_{\rm tot,z}<14\msun$) contain predominantly 
BH-NS systems. Then higher mass bins are dominated by BH-BH mergers. This will be 
demonstrated below (see Sec.~\ref{results2} and Fig.~\ref{f6}). 

In models V1, V2, and V4 the merger rate density increases for these bins up to
a total merger mass of $\sim 30\msun$. At higher masses the merger rate density is 
steadily declining, but massive binary mergers are still predicted in all of our 
models up to a total redshifted mass of $\sim 120\msun$. 

Predictions for the model with high BH kicks (V3) show a different behavior than other 
models (V1, V2, V4). There is a rather flat merger rate density dependence on total merger 
mass for a broad spectrum of total redshifted masses ($10$--$110\msun$). The dip for the 
lowest mass BH-NS bin ($M_{\rm tot,z}=5$--$8\msun$) is significantly more pronounced 
than in other models. These changes are naturally explained by high natal kicks applied 
to all BHs independent of their formation mass. The high natal kicks tend to disrupt 
potential progenitors of BH-BH and BH-NS mergers.   

Predictions for the delayed SN engine model (V4) closely resemble the standard model 
(V2). The only notable difference comes in the second mass bin ($M_{\rm tot,z}=5$--$8\msun$) 
in which the V4 model generates a merger rate density almost as high as for our most 
optimistic model V1 (see Table~\ref{t3} and ~\ref{t4}). This relatively high merger rate 
density is produced by low mass mergers (either BH-BH or BH-NS) with compact objects with 
mass in the mass gap range ($2-5 \msun$) that are naturally formed within the evolutionary 
framework of model V4. We do not show this model to avoid overcrowding on the plots. 

Generally, the merger rate density decreases from model V1 to V2 to V3. For example, for low 
metallicity evolution the merger rate density drops by a factor of a $\sim$ few from V1 to V2,  
and then down by $\sim$ one--two orders of magnitude from model V2 to V3. These changes are 
very similar for high metallicity evolution (see Table~\ref{t4}). The decreasing merger rate 
density progression is easily understood in terms of the  input physics associated with our 
evolutionary models. Model V1 allows for very optimistic evolution in the context of double 
compact object formation (in addition to CE during core Helium burning, case B RLOF evolution 
allows the formation of close post-CE binaries). In model V2 the formation routes for close 
double compact objects are limited to case C RLOF (CE only during core Helium burning) and 
thus the merger rate densities decrease. Finally, the high BH kicks present in model V3 
severely reduce the rate density for BH-BH and BH-NS mergers. 

Model V1 merger rate densities for low metallicity evolution are the highest within our 
sample of models. In particular, the predicted merger rate densities are only a factor of 
$18$ below the initial LIGO/Virgo upper limits in the $M_{\rm tot,z}=25$--$37\msun$ bin. The 
more realistic model V2 results in a BH-BH merger rate density a factor of $\sim 70$ below 
the initial LIGO/Virgo upper limits for a total merger mass of $M_{\rm tot,z}=25$--$54\msun$.
For comparison, note that the predicted merger rate density in the lowest mass bin (NS-NS 
mergers) is a factor of $\sim 1000$ below the upper limits for all models and for both 
metallicity evolution scenarios. This indicates that BH-BH mergers are expected to be 
dominant GW sources within the framework of the evolution adopted in models V1, V2, and V4. 
If no detections are made after a modest increase of LIGO/Virgo sensitivity, these 
specific evolutionary models will be eliminated in the context of BH-BH formation. 

Our calculations incorporate the evolution of average metallicity with redshift into 
the cosmological model of the Universe; a broad spectrum of metallicities is employed in our 
population synthesis rate predictions. Changing the details of the adopted metallicity 
evolution model, within the uncertainties, produces at best modest effects on the rate 
predictions. Specifically, the low metallicity models produce somewhat higher merger rate 
densities for heavy mergers (BH-BH) and somewhat lower merger rate density for low-mass 
mergers (NS-NS). These trends are understood in terms of the effects of metallicity on the BH-BH 
formation (low metallicity boosts the formation) and on the NS-NS formation (low metallicity 
suppresses the formation; see Dominik et al. 2012).
For example, the standard model (V2) merger rate density changes from $1.3\times10^{-8}\mpy$ 
for low metallicity evolution (see Table~\ref{t3} and Fig.~\ref{f2}) to $7.0\times10^{-9}\mpy$ 
for high metallicity evolution (see Table~\ref{t4} and Fig.~\ref{f3}) for BH-BH mergers with 
total mass $M_{\rm tot,z}=25$--$37\msun$. The decrease of merger rate density from low to high 
metallicity evolution is rather moderate: only a factor of $\sim 2$. 

Note that metallicity itself plays a crucial role in the formation process of BH-BH binaries 
(Belczynski et al. 2007; Belczynski et al. 2010a). For example, a group of stars with
solar metallicity will form $\sim 10$--$100$ times fewer BH-BH mergers than the same group 
with $10\%$ solar metallicity (de Mink \& Belczynski 2015). However, once we incorporate this 
very strong dependence into cosmological models of the Universe and employ a broad spectrum of
metallicities (as in our methods II and III), the merger rate density predictions are not as 
strongly affected. Both metallicity evolution models generate (alas with a different efficiency) 
low metallicity stars (typical progenitors of BH-BH mergers).  A broad BH-BH merger delay time 
distribution (Dominik et al. 2013, 2015) allows for both models to produce BH-BH binaries that 
will merge at low redshifts.

\subsection{Comparison of Models with Forecasted Advanced LIGO/Virgo Upper Limits}
\label{results2}

In Figures~\ref{f4} and~\ref{f5} and Tables~\ref{t5} and~\ref{t6} we present 
evolutionary model predictions for double compact object merger rate densities. 
The predictions are constructed with method II (see Sec.~\ref{method2}) and 
contrasted with the forecasted advanced LIGO/Virgo upper limits (see Sec.~\ref{aligo}). 
Our models are described in Sec.~\ref{models}. 

The predicted double compact object merger rate density within the volume sampled by the 
advanced LIGO/Virgo detectors are high enough relative to the forecasted upper limits to 
make detections very likely for most of our considered evolutionary models: V1, V2, and V4 
for both metallicity evolution scenarios. Model V4, which is not shown in the figures, is 
very similar to model V2 (and V4 merger rate densities are given in the tables). Predictions 
for model V3 are well below the upper limits in all mass bins, with the exception of the NS-NS 
mass bin ($2$--$5\msun$). This does not make detections impossible, it just indicates that the 
detections are highly unlikely for this evolutionary scenario. The difference between 
predictions and the upper limits is a gauge of how likely (or unlikely) the detections 
will be. In the case of models V1, V2 and V4 the detections are highly likely, as merger 
rate densities are well above the upper limits in some mass bins. If there is no detection, 
these models can be reliably excluded as incorrect.   
For our standard input physics and low-metallicity evolution (model V2l), the three mass 
bins that are the highest above the advanced LIGO/Virgo upper limits are: 
$M_{\rm tot,z}=25$--$37\msun$, with merger rate density is $5.0$ times above upper limits; 
$M_{\rm tot,z}=37$--$54\msun$, with merger rate density is $12.5$ times above upper limits;
$M_{\rm tot,z}=54$--$73\msun$, with  merger rate density is $6.4$ times above upper limits
(see Table~\ref{t5}). This indicates the most likely mass range for future detections. 
This is a generic feature of our predictions found in most models (V1, V2, V4) for 
both metallicity evolution scenarios.
The only exception is model V3 with high BH natal kicks, with merger rate
density consistently below 
the upper limits. Note that our comparison is done for one year of observations (that 
is how upper limits are calculated). The advanced LIGO/Virgo detectors are scheduled to 
work for approximately $\sim 10$ years at full sensitivity. However, even if we made the 
upper limits better by a factor of $10$, model V3 merger rate densities for BH-NS and 
BH-BH mergers would still end up below the improved $10$ year upper limits. 

Note that our predictions are close to the upper limits for all the models of NS-NS 
mergers (the first mass bin). This makes detections questionable, but not impossible, 
within $1$ year of observation, but quite likely within $10$ years.

The results for the advanced detectors may be contrasted with the predictions for
initial LIGO/Virgo: all the models are below the upper limits. This difference comes from 
two facts. 
First, the advanced instruments will be much more sensitive (by a factor of
$\sim 10$), so the upper limits (per unit of observation time) will be associated with 
much larger searched  volume ($\sim 1000$) than initial LIGO/Virgo.
Second, the predicted merger rate densities within the volumes sampled by the advanced 
instruments are generally higher than those predicted for volumes sampled by the initial 
instruments. Initially, merger rate densities increase with increasing 
redshift to $z=1$--$4$ (depending on the type of merger), and then they decline 
with redshift (Dominik et al. 2013; see their Fig.~3 and 5). This is a result of
the merger rate density approximately following star formation rate (that peaks at 
$z \approx 2$) and the contribution of mergers that are formed at high redshifts in 
low metallicity stellar populations with long delay times. This effect becomes more 
pronounced with total merger mass as more massive mergers can be detected from higher 
redshifts. In particular, for our standard model V2 and low metallicity evolution, the 
difference in merger rate density is negligible for NS-NS mergers ($5.0\times10^{-8}\mpy$ 
for initial LIGO/Virgo versus $5.7\times10^{-8}\mpy$ for advanced LIGO/Virgo); it becomes 
more pronounced for $M_{\rm tot,z}=37$--$54\msun$ ($4.5\times10^{-9}\mpy$ for initial 
LIGO/Virgo versus $1.5\times10^{-8}\mpy$ for advanced LIGO/Virgo); and it is significant 
for higher mass bins (e.g., for $M_{\rm tot,z}=91$--$109\msun$ it is  $1.6\times10^{-11}\mpy$ 
for initial LIGO/Virgo versus $4.5\times10^{-10}\mpy$ for advanced LIGO/Virgo). 

We also note that all our model predictions extend to very high total redshifted 
merger mass: $\sim 400\msun$ for models V1, V2, V4 and $\sim 350\msun$ for model V3.
We have already explained this surprising finding (Sec.~\ref{methods}) in the context of 
model V2. A very low metallicity progenitor ($Z=0.0001$) binary with very high initial
component masses ($m_{\rm zams,1}=148\msun$ and $m_{\rm zams,2}=144\msun$) forms a massive 
BH-BH system (intrinsic total BH-BH mass of $136\msun$) that merges at redshift $z=2$ (total 
redshifted mass of $\sim 400\msun$). Such a merger is potentially detectable at full advanced 
LIGO/Virgo sensitivity with the use of the full waveform  (inspiral-merger-ringdown).
Models V1 and V4 show the same final outcome. For model V3 high BH natal kicks disrupt 
a large number of BH-BH progenitors and the most massive binaries (with associated small 
number statistics) are not found in our Monte Carlo population synthesis simulations.

BH-NS mergers provide only a relatively small contribution to high mass bins for models V1, 
V2 and V3 and moderate contribution to high mass bins for model V4 (see ~\ref{f6}). 
BH-NS mergers occupy mass bins in the range $M_{\rm tot,z}=5$--$73\msun$
for models V1, V2, V3 and $M_{\rm tot,z}=2$--$127\msun$ for model V4. The highest merger 
rate densities for these mergers are found in the mass range $M_{\rm tot,z}=8$--$17\msun$
for models V1, V2, V3. For model V4 the highest three bins include total redshifted merger 
mass in the range $M_{\rm tot,z}=2$--$5\msun$ and $M_{\rm tot,z}=25$--$54\msun$. Both our 
optimistic model (V1) and delayed SN engine model generate merger rate densities that are 
at the level of (or higher than) the upper limits ($1$ yr of observations). The standard 
model (V2) merger rate densities for BH-NS binaries are about an order of magnitude below 
the upper limits, making detections unlikely during $1$ yr and likely during $10$ yr of 
observations. Model V3 merger rate densities are quite low and make BH-NS detections 
unlikely. 

In summary: the predictions, within optimistic-to-realistic models, show likely detections 
for NS-NS mergers (first mass bin), possible detections for BH-NS mergers (second and higher 
mass bins), and very likely detections for BH-BH mergers (third and higher mass bins) at the
full advanced LIGO/Virgo sensitivity. Pessimistic model indicates likely detections for
NS-NS mergers and unlikely detections for BH-NS and BH-BH mergers.

\section{Black Hole Natal Kicks}
\label{BHkicks}

We have compiled the empirical and theoretical information on BH natal kicks.
The empirical data is presented in Table~\ref{T:kicks} and Figure~\ref{bhkick}
and discussed in Appendix~\ref{Appkicks}. The comparison of empirical estimates with our models
and discussion of physics behind natal kicks is given in the following two
subsections.

\subsection{Comparison of Empirical Estimates with Models}
\label{compkicks}

In this section we will contrast the empirically derived information on BH masses and natal 
kicks with population synthesis models of Galactic BH interacting binaries. 

In one approach to BH natal kicks in core collapse SNe we employ the modified (decreased) 
Maxwellian kick distribution with $\sigma=265\kms$.  The mass of the 
remnant object generates  gravitational potential strong enough to prevent parts or all 
of the mass ejected during SN from reaching escape velocity. The matter falling back onto the 
remnant object will reduce the original (Maxwellian) kick velocity due to conservation of 
momentum. The more massive the final (pre-SN) core of the star, the more fallback it generates.
This results in BH natal kicks smaller than the NS kicks. To account for this effect we use a 
simple linear relation for the reduction of the  natal kick magnitude by the amount of fallback 
during a SN:
\begin{equation} \label{vkick}
V_k=V_{max}(1-f_{\rm fb}),
\end{equation}
where $V_k$ is the final magnitude of the natal kick, $V_{max}$ is the velocity drawn from a 
Maxwellian kick distribution, and $f_{\rm fb}$ is the fallback factor. The values of 
$f_{\rm fb}$ range between $0$--$1$, with $0$ indicating no fallback/full natal kick and $1$ 
representing total fallback/no kick (direct BH formation without any SN event and without any 
mass loss). Additionally to the natal kick, the  Blaauw kick is calculated from symmetric 
mass ejection (if any).  The combination of two kicks may disrupt the binary, or  generate a 
peculiar systemic velocity of the surviving binary in its motion in Galaxy. The values for the 
fall back and BH masses for our models are introduced and described in Fryer et al. (2012).
This approach to natal kicks is adopted in our models V1,V2 and V4 and we refer to this 
scenario as ``low BH kicks''. This method generates a trend of natal kicks decreasing with BH 
mass. In particular massive BHs do not receive any natal kick. 

In another approach to BH natal kicks we draw a Maxwellian 1D velocity distribution with 
$\sigma=265\kms$ independent of BH mass.
This distribution of natal kicks is based on the observed velocities of single Galactic 
pulsars (Hobbs et al. 2005). Such a distribution allows for very high BH natal kicks
with the averaged 3D natal kick magnitude of $420\kms$. The kick direction is assumed 
to be random. We refer to this scenario as ``high BH kicks'' and employ this scheme in our 
evolutionary model V3. Note that this model allows for low natal kicks. In particular, 
our evolutionary model will eliminate the highest kicks drawn from this distribution by 
disrupting binary stars. 

In Figure~\ref{bhkick} we show empirical estimates of the BH natal kicks and BH masses. 
Both the BH natal kicks and the mass estimates suffer large uncertainties. Along with the
empirical estimates we show two theoretical evolutionary predictions of Galactic 
population of interacting BH binaries. The two models differ only in one assumption; 
either low or high BH kicks are adopted.  The models are calculated for stars with solar 
metallicity ($Z=0.02$) and with our standard evolutionary model (see Sec.~\ref{models}). 
We stop evolution at the point of BH formation. We choose only binaries with 
orbital separations smaller than $100\rsun$. The longest period BH binary in our empirical 
sample (GRS 1915+105) has period of $P_{\rm orb}=33.5$d (Greiner, Cuby \& McCaughrean 2001) 
that corresponds to the orbital separation of $\sim 100\rsun$. We then assume that each of 
these binaries will eventually undergo RLOF and become an X-ray binary. We do not model 
RLOF phase (there seem to be a general issue with Galactic BH binary modeling; i.e., 
Wiktorowicz et al. 2014). Instead we assume that some fraction (randomly chosen value 
between $0$ and $0.2$) of the donor star can accrete onto a BH. We plot the BH masses
increased in this process versus the natal kicks these BHs have been assigned in our 
``low-kick'' or ``high-kick'' scenario (Fig.~\ref{bhkick}). In this approach we have not 
taken into account any potential observational biases as we lack a good model for X-ray 
emission for BH systems. This is in particular true for BH transients that make most of 
our sample (all, but Cyg X-1, systems in Table~\ref{T:kicks} are transients).  

Note that our evolutionary models take into account various binary configurations at the time 
of BH formation, and account for survival and disruptions of systems on arbitrary orbits. Only 
systems that are tightly bound or those that receive small or preferably oriented natal kicks 
survive BH formation. We note a number of very small (or no) natal kick binaries and with kicks 
generally below $300\kms$ for our low BH kick model. For high BH kick model, the kicks in 
surviving binaries are non-negligible, with most found in range $50$--$500$ km s$^{-1}$.   
As clearly seen from Figure~\ref{bhkick}, while neither of the models provides a good match to 
the empirical estimates, both models are consistent with the empirical estimates (i.e., we can 
find synthetic binaries nearby each empirical point for both natal kick models). On one hand 
this reflects the fact that the empirical data is still very poor (only 5 good estimates and 
many weak lower limits). On the other this demonstrates that even low-kick model may deliver 
a range of BH natal kicks that are consistent with the empirical estimates. For low-mass BHs 
there is significant mass ejection so both low and high kicks are expected depending on the 
level of asymmetry. For high mass BHs, low to zero kicks are expected if a BH forms with high 
mass (small to no mass ejection) and higher kicks are expected for a BH that has formed at 
low mass and then increased its mass via accretion from its companion. 
For high BH kick model (despite its name) low kicks at the level of $\sim50\kms$ are predicted 
in our evolutionary simulations for some of the interacting BH binaries since BH natal kicks 
are drawn from the Maxwellian distribution. 

Contrary to some common beliefs that natal kicks decrease with BH mass, we point out that 
both theoretical models: natal kicks independent of BH mass (our high-kick model) and kicks 
decreasing with BH mass (our low-kick model), 
can explain the empirical data within their associated errors. In particular, Mirabel \& 
Rodrigues (2003) and Dhawan et al. (2007) claim that the most massive BHs in Galaxy (GRS 
1915+105; $M_{\rm BH}=12.4\msun$ and Cyg X-1; $M_{\rm BH}=14.8\msun$) form at dark (without 
a supernova) and without any significant natal kick, while lower mass BHs (e.g., XTE J1118+480; 
$M_{\rm BH}=7.6\msun$) form at supernova with significant mass ejection asymmetry and 
associated high natal kick. This is at first glance very convincing, however other systems may 
challenge such notion. For example, GS 2023+338 with $M_{\rm BH}=9.0_{-0.6}^{+0.2}\msun$ has 
very small natal kick $V_{\rm kick}<45\kms$, while more massive BH in XTE J1819-254 
($M_{\rm BH}=10.2\pm1.5\msun$) has significant natal kick $V_{\rm kick}>100\kms$. Apparently, 
this counter-example may go away if errors on BH mass estimates are allowed to work in favor 
of Mirabel \& Rodrigues (2003) and Dhawan et al. (2007) claims. If one applied similar type of 
arguments on allowed errors on natal kicks, it may be claimed that the second lowest mass BH 
on our list: GRO J1655-40 ($M_{\rm BH}=5.3\pm0.3\msun$) formed with no natal kick, while the 
two heaviest BHs (in Cyg X1 and GRS 1915+105) have formed with kicks in excess of $50\kms$ 
(see Table~\ref{T:kicks} for actual estimates and their associated errors). 

At this point, we prefer to stay agnostic, allowing for both natal kick models: independent of 
BH mass and decreasing with BH mass. Further precise determinations of BH masses and BH natal 
kicks, are in our opinion needed to either support or reject claims that natal kicks decrease 
with BH mass.

\subsection{Natal Kick Physical Mechanisms}
\label{kicks}

Neutron star and black hole binaries can receive peculiar velocities in systems where the 
compact object forms in a supernova explosion with associated symmetric mass ejection. The 
binary system moves in the opposite direction of the ejecta (Blaauw 1961). This mechanism is 
not sufficient to explain the observed pulsar velocity distribution and a number of additional 
"natal kick" mechanisms have been proposed to explain the proper motions of NSs. These kick 
mechanisms have implications for BH natal kicks and we will review them here. Natal kicks 
occur when there are asymmetries in either the matter ejected or the neutrinos emitted 
during the core collapse and/or supernova explosion. The  velocity of the newly formed 
compact object is determined by the momentum conservation.

Normal supernovae are believed to be driven by the convection-enhanced engine where 
convection between the proto-NS and the edge of the stalled shock plays an important role 
in increasing the efficiency at which the gravitational potential energy is converted into 
energy driving an explosion. Both theoretical and observational evidence for this explosion 
engine as a mechanism behind normal supernovae has grown (for a review, see Fryer et al. 
2014). Almost immediately after realizing that this convection could be important for the 
explosion, it was realized that low-mode convection could drive natal kicks (Herant 1995).  
Since this time, a series of calculations have been conducted to study the kicks from these 
convection cells (Scheck et al. 2006; Fryer \& Young 2007). Although models have been 
constructed to obtain extremely high kicks, many models predict kicks in the $100$--$200\kms$ 
range. Alternatively, asymmetric ejecta can be driven by asymmetries in the density profile 
of collapsing core of a massive star (Burrows \& Hayes 1996; Fryer 2004).

Although there is no decided ejecta kick mechanism, we can still make predictions for BH 
kicks assuming such mechanisms work. Momentum conservation sets the momentum of the BH 
(${\bf p}_{\rm BH}$) to be equal (but in opposite direction) to the momentum of the ejecta 
(${\bf p}_{\rm ej}$): 
\begin{equation}
{\bf p}_{\rm BH} = -{\bf p}_{\rm ej}.  
\end{equation} 
If the ejected material has a fixed asymmetry ($\alpha_{\rm ej}$), then the BH kick velocity 
($v_{\rm BH}$) is
\begin{equation}
v_{\rm BH} = {\alpha_{\rm ej} M_{\rm ej} \over M_{\rm BH}} <v_{\rm ej}> = 
             {\alpha_{\rm ej} M_{\rm ej} \over M_{\rm star}-M_{\rm ej}} <v_{\rm ej}>
\label{eq:vkickej1}
\end{equation}
where $M_{\rm ej}$ is the ejecta mass, $M_{\rm BH}$ is the black hole mass, $M_{\rm star}$ 
is the mass of the star at collapse and $<v_{\rm ej}>$ is the mass averaged velocity of the 
ejecta. In this formulation, if $\alpha_{\rm ej}$ is constant, the kick velocity decreases 
with decreasing ejecta mass. Or to put it differently, asymmetric natal kick decreases with 
increasing BH mass. But, in the convective engine, the asymmetry in the explosion grows with 
time as the convection approaches lower modes. Late-time explosions are believed to be less 
energetic (with less ejecta and more massive BHs). In this scenario, the ejecta asymmetry 
increases with the explosion delay (and, hence, BH mass) and the BH kick can increase with 
increasing BH mass. Depending on the ejecta asymmetry behavior the magnitude of the natal 
kick may increase (fixed $\alpha_{\rm ej}$) or decrease ($\alpha_{\rm ej}$ increasing with SN 
delay) with BH mass. 

If the kick is instead produced by asymmetries in the pre-collapse stellar structure, the 
asymmetry is strongest at bounce and decreases with explosion delay. It means that the natal 
kick will decrease with increasing BH mass even faster than predicted by equation~\ref{eq:vkickej1}. 
Without understanding which asymmetry drives natal kick and without knowledge of the evolution 
of the asymmetry with time, it is difficult to predict exact trends.

Asymmetries in the neutrino emission can be caused by a variety of mechanisms. Initial models 
focused on active ($e, \mu, \tau$) neutrinos, showing that these neutrinos can be produced with 
high asymmetries (Bisnovatyi-Kogan 1993; Lai \& Qian 1998; Arras \& Lai 1999; Kei, Shoichi \& 
Katsuhiko 2005). Although these neutrinos can be produced with high asymmetries ($\sim 30\%$), 
by the time these neutrinos reach the neutrinosphere and escape, they are much more isotropic.  
If sterile neutrinos are produced, their immediate escape means that the escaping neutrinos are 
also asymmetric (Kusenko \& Segre 1997; Nardi \& Zuluaga 2001; Fuller et al. 2003; Kusenko 2005; 
Fryer \& Kusenko 2006; Sagert \& Schaffner-Bielich 2008; Kusenko 2009; Kisslinger, Henley \& 
Johnson 2009; Kuznetsov \& Mikheev 2012). For these sterile neutrinos, the magnitude of the kick 
depends also upon the fraction of energy converted into sterile neutrinos. Alternatively, 
neutrino asymmetries can be produced through radiatively-driven magnetoacoustic instabilities at 
the neutrinosphere (Socrates et al. 2005). Neutrino asymmetries can occur even without strong 
magnetic fields if asymmetric convection can induce asymmetries in the neutrino emission. The 
kick magnitudes from these asymmetries are typically low (for a review, see Janka 2013).

As with the ejecta kick mechanisms, it is difficult to estimate the natal kick dependence on BH 
mass without using a specific model. The collapse of a massive star releases a collapse energy: 
\begin{equation}
E_{\rm coll}={G M^2_{\rm NS} \over r_{\rm NS}} \approx 2.7\times10^{53}
             \left({M_{\rm NS} \over \msun}\right)^2 \ {\rm erg},
\label{ecoll}
\end{equation}
most of which is carried away by neutrinos (with a momentum of $E_{\rm coll}/c$ where $c$ is the 
speed of light).  If we assume the neutrino asymmetry is constant until the core collapses to a 
BH, the kick produced by asymmetric neutrino emission is:
\begin{equation}
v_{\rm BH} = 150 {\alpha_\nu \over 0.01} {3\msun \over M_{\rm BH}} 
                   {E_{\rm coll} \over 2.7\times10^{53} \ {\rm erg}}
\label{eq:nukick1}
\end{equation}
where $\alpha_\nu$ is the asymmetry in the neutrinos. This relation generates natal kicks that 
decrease with BH mass. However, it also indicates that even massive BHs may receive significant 
natal kicks. For example, if we assume that the maximum NS mass is $2.2\msun$, the average 
asymmetry of the neutrinos is $1\%$, a BH of mass $10\msun$ could receive a kick in excess of 
$200\kms$. 

If the stellar core is not rotating rapidly, the neutrino emission will drop dramatically after 
the neutron star collapses to a black hole. Any neutrino kick mechanism will shut down at the 
formation of the black hole (this is assumed in our estimate of the available collapse energy;
eq.~\ref{ecoll}). Even if the magnetic field grows with time, producing a stronger neutrino 
asymmetry with time, this turn-off will likely produce weaker kicks for more massive black holes. 
However, if the core is rotating rapidly, neutrino emission from the black hole accretion disk 
can dominate the total neutrino budget (e.g., neutrinos from collapsar model: Popham, Woosley \& 
Fryer 1999; Fryer \& Meszaros 2003). If some mechanism can be devised to produce asymmetric 
emission from this disk (e.g. instabilities similar to those discussed by Socrates et al. 2005), 
the BH kick could grow with increasing black hole mass.  

Unfortunately, both the ejecta and neutrino mechanisms predict the same trends in general: more 
massive black holes will have lower kicks. There are some differences. Most of the neutrino 
mechanisms argue for kicks aligned with the spin axis and this need not be the case for the 
ejecta mechanisms (although, the ejecta velocity is likely to be aligned with the spin axis in 
rapidly rotating systems).  In neutrino mechanisms, the kick need not be directed in the 
opposite direction of the ejecta. Indeed, for some mechanisms, the ejecta and kick can be in the 
same direction (Fryer \& Kusenko 2006). Finally, the neutrino-driven kick mechanism can occur 
even if there is no supernova explosion (no mass loss/ejecta) at all.
Although measurements of natal kicks and their dependence on BH mass may not point  
unambiguously to a particular kick mechanism, they could provide very useful information on 
development of asymmetry in core collapse/supernova explosion. Whether the asymmetry comes 
in the form of mass ejection or neutrino emission, such measurements would allow to 
constrain and improve supernova models.

\section{Potential Reduction of BH-BH Merger Rates}
\label{nobhbh}

In case of the non-detection of massive BH-BH mergers, high black hole natal kicks (as 
we argue throughout this study) are one of two current proposals for how to limit their 
formation. Mennekens \& Vanbeveren (2014) find no BH-BH mergers in their population 
synthesis predictions. This finding applies to BH-BH mergers of any mass and it is 
connected with very intensive stellar wind mass loss. Intensive mass loss not only 
decreases the amount of mass available for BH formation, but also limits the radial 
evolution of a star. The expansion is required at late evolutionary stages to lead to 
RLOF that needs to develop into CE, allowing for orbital contraction and the formation 
of a coalescing BH-BH binary. Mennekens \& Vanbeveren (2014) claim that the mass loss 
for BH progenitors may be so intensive (especially during the LBV phase) that it bars 
BH-BH formation.  

In general, the radial expansion of massive stars is not fully understood. Radial expansion 
may be limited on the outside by strong stellar winds removing the H-rich envelope, and on 
the inside by mixing envelope H-rich material into the core (either by strong convection 
and/or by rapid rotation). 

At high metallicity (strong stellar winds) it is claimed that 
massive stars above $M_{\rm zams}>40\msun$ do not significantly expand as they are not 
observed in specific parts of the H-R diagram (e.g., Mennekens \& Vanbeveren 2014).
However, stars with somewhat lower mass $\sim 20$--$40\msun$ are observed to expand to 
large radii (AH Sco: $1400\rsun$, UY Sct: $1700\rsun$, KW Sgr: $1000\rsun$) in the Milky 
Way, so at rather high metallicity (Arroyo-Torres et al. 2013). 

At low metallicity (i.e. $10\%$ solar, at and below which we predict effective 
close BH-BH formation) there is no available observational information on the radii of 
massive stars. Evolutionary models show that already at SMC metallicity ($20$--$30\%$ solar) 
stars up to $60\msun$ expand to large radii and become red super giants (Brott et al. 2011). 

Although the mass range for BH formation is not well constrained, with claims that change 
from a lower limit for BH formation of $M_{\rm zams}>20\msun$ (e.g., Dominik et al. 2014) 
to $M_{\rm zams}>40\msun$ (e.g., Mennekens \& Vanbeveren 2014) to no limit at all with 
NS/BH formation being a rather chaotic function of initial star mass (Clausen, Piro \& Ott 
2015) there seems to be still a lot of parameter space available for radial expansion of 
potential BH progenitors. 

Stellar winds tend to increase orbital separation between two massive stars in a binary 
system. If wind mass loss is large (as expected for massive stars) and if stars do not 
expand significantly, binary components may never reach RLOF/CE. Tidal interactions may be 
a potential mediating factor. If a massive star expands after main sequence it slows down, 
if it is not already a slow rotator. If a star radius reaches about half the size of its 
Roche lobe the tides will tend to spin the star up at the expense of the binary angular 
momentum. This effect may be stronger than the increase of orbital separation due to wind 
mass loss. The orbit decreases and the star may initiate RLOF/CE. Such a case was 
demonstrated for a $40\msun$ star with the extreme LBV mass loss adopted from Mennekens \& 
Vanbeveren (2014). This star was placed in a binary with a $7\msun$ black hole, 
and the binary was shown to form a typical BH-BH merger despite the intense wind mass loss 
rate (Dominik et al. 2015). On the other hand, it is unclear how effectively tides 
dissipate energy. If tidal energy dissipation is not very effective the orbital 
contraction (and thus RLOF/CE) is not expected unless a star does not expand much on its 
own, as in such a case strong winds dominate the evolution of binary separation. Tidal 
interactions in close binary systems are not taken into account in the evolutionary model 
of Mennekens \& Vanbeveren (2014), so they have not considered the effects of this 
mitigating factor on their no BH-BH merger proposal.

\section{NS-NS merger rates}
\label{nsns}

Our predicted rate densities for NS-NS mergers for advanced LIGO/Virgo are presented in 
Figures~\ref{f4} and~\ref{f5} and Tables~\ref{t5} and~\ref{t6} (the first mass bin). 
These merger rate densities, although low ($\sim 0.5$--$1.7 \times 10^{-7}\mpy$), are 
consistent with available empirical estimates. 

Kim et al. (2010) have estimated NS-NS merger rate based on 
observations of three Galactic field NS-NS systems, B1913+16, B1534+12, and J0737-3039, 
and found a Galactic  merger rate within the range $3$--$190$ Myr$^{-1}$. O'Shaughnessy 
\& Kim (2010) obtained a median value of $89$ Myr$^{-1}$, with a spread above and below 
by a factor of $\sim 3$ when pulsar beaming constraints are taken into account. Kim, 
Perera \& McLaughlin (2015) have re-examined the influence of double pulsar 
J0737-3039\footnote{Note that it's not the double pulsar that decreases the previous rate 
estimates}, obtaining a revised estimate of the Galactic merger rate of 
$7$--$49$ Myr$^{-1}$, with median value $21$ Myr$^{-1}$. These headline numbers do 
not include the large uncertainties in the pulsar luminosity function. If these 
uncertainties are included, it is expected that the rates could shift up or down by an 
order of magnitude (Kalogera et al. 2004; Mandel \& O'Shaughnessy 2010; Abadie et al. 
2010). Applying this to the most recent estimate results in a broad range of allowed 
rates: $2.1$--$210$ Myr$^{-1}$. 

For comparison, the Galactic merger rates\footnote{These rates are obtained with the 
assumption of Galactic $10$ Gyr of constant star formation rate at the level $3.5\msun$ 
yr$^{-1}$.} in our standard and optimistic evolutionary scenario are $7.6$ and $23.5$ 
Myr$^{-1}$, respectively. Note that these {\em Galactic} rates are not listed separately 
in any of our tables, but are used for all the predictions, 
contributing a part of the over-all NS-NS merger rate (which results from galaxies of 
various metallicities). The actual Galactic merger rates (that are used here) are reported 
in Dominik et al. 2012; see their Table 2 for $Z=0.02$). These rates are relevant for 
Galactic field evolution where the majority of NS-NS binaries are found. Our NS-NS merger 
rates for sub-solar metallicity ($Z=0.002$; Table 3 of Dominik et al. 2012) are a factor of 
$\sim 3$ lower. The full spread of NS-NS Galactic rates for the solar metallicity models is 
$23.3$--$77.4$ Myr$^{-1}$ for the Dominik et al. (2012) A submodels (they correspond to our 
optimistic assumption on CE) and $0.3$--$9.5$ Myr$^{-1}$ for the B submodels (our standard 
model assumption on CE). The predicted NS-NS merger rates are fully consistent with Galactic 
NS-NS observations for our optimistic model (V1), and are on the lower end of empirical 
estimates for our other models (V2, V3, V4). In particular, our simple estimate is consistent 
with the detailed analysis of pulsar luminosity function uncertainty by O'Shaughnessy \& Kim 
(2010; see their Fig 11).

There is mounting evidence that short Gamma-ray bursts (GRBs) are connected with NS-NS and/or 
BH-NS mergers (e.g., Berger 2013), and some authors have used short GRB rates to estimate NS-NS 
merger rates (e.g., Chen \& Holz 2013). Fong et al. (2012) found short GRB rate densities at 
the level $100$--$1,000$ Gpc$^{-3}$ yr$^{-1}$, while Enrico Petrillo, Dietz, \& Cavaglia (2013) 
estimated the rate density to be $500$--$1,500$ Gpc$^{-3}$ yr$^{-1}$. These results suffer 
from short GRB beaming and luminosity uncertainties. The beaming has been firmly established 
for $\sim 3$ short GRBs (with $\theta_j\lesssim 10\,\deg$), while redshifts and thus 
luminosities are only known for the $\sim 20$ closest events. It is possible that the average 
beaming angle is larger than currently estimated, and that the rates densities are 
correspondingly lower, possibly by an order of magnitude (E.Berger 2013, private 
communication). The lower limit on the short GRB rate density would then decrease to $10$ 
Gpc$^{-3}$ yr$^{-1}$. Before comparing with our NS-NS merger rate density it is worth noting 
that even if NS-NS mergers are in fact short GRB progenitors, they may be responsible for only 
a fraction of short GRBs as other progenitors cannot be excluded at the moment (e.g., Nakar 
2010). On the other hand only some fraction of NS-NS mergers may produce short GRBs (e.g., 
Fryer et al. 2015). The NS-NS merger rate densities that we report here (tables and figures) 
are at the level of $50$--$150$ Gpc$^{-3}$ yr$^{-1}$. We note that our reported merger rate 
densities apply {\em only}\/ to the local Universe within the reach of advanced LIGO/Virgo for 
NS-NS mergers ($450$ Mpc; $z<0.1$). Our NS-NS merger rate densities are predicted to 
moderately increase with redshift at low redshifts (see Fig.~3 in Dominik et al. 2013). 
Considering all of the above, our reported NS-NS merger rate densities are only a lower limit 
on the overall GRB rate density, and thus they are consistent with GRB rate predictions. 

Finally, our merger rate densities for NS-NS mergers ($50$--$150$ Gpc$^{-3}$ yr$^{-1}$) are 
consistent with kilonova rate estimated from the most recent event reported to accompany a
long-short GRB 060614: $\gtrsim 10$ Gpc$^{-3}$ yr$^{-1}$ (Jin et al. 2015).
If the event was generated by BH-NS merger then such event would be consistent only with some 
of our models. Merger rate density for BH-NS binaries is at the level $\sim 10$-$100$ Gpc$^{-3}$ 
yr$^{-1}$ for our optimistic and delayed SN models, and at the level of $\sim 1$ Gpc$^{-3}$ 
yr$^{-1}$ for our realistic and high BH kick models (see Fig.~\ref{f6}).

\section{Conclusions}
\label{conc}

{\em (I) Executive summary.}
We performed a comparison of evolutionary predictions for NS-NS, BH-NS, and BH-BH mergers 
with observational upper limits for gravitational wave detectors. We note that, due to 
cosmological and evolutionary effects, some care must be taken in the process of making 
these comparisons. In particular, in the case of advanced LIGO/Virgo detectors with their 
significant detection horizon (reaching redshift of $z=2$ for BH-BH mergers), a careful 
analysis of evolutionary and population synthesis predictions is required.

\begin{figure}
\includegraphics[width=1.0\columnwidth]{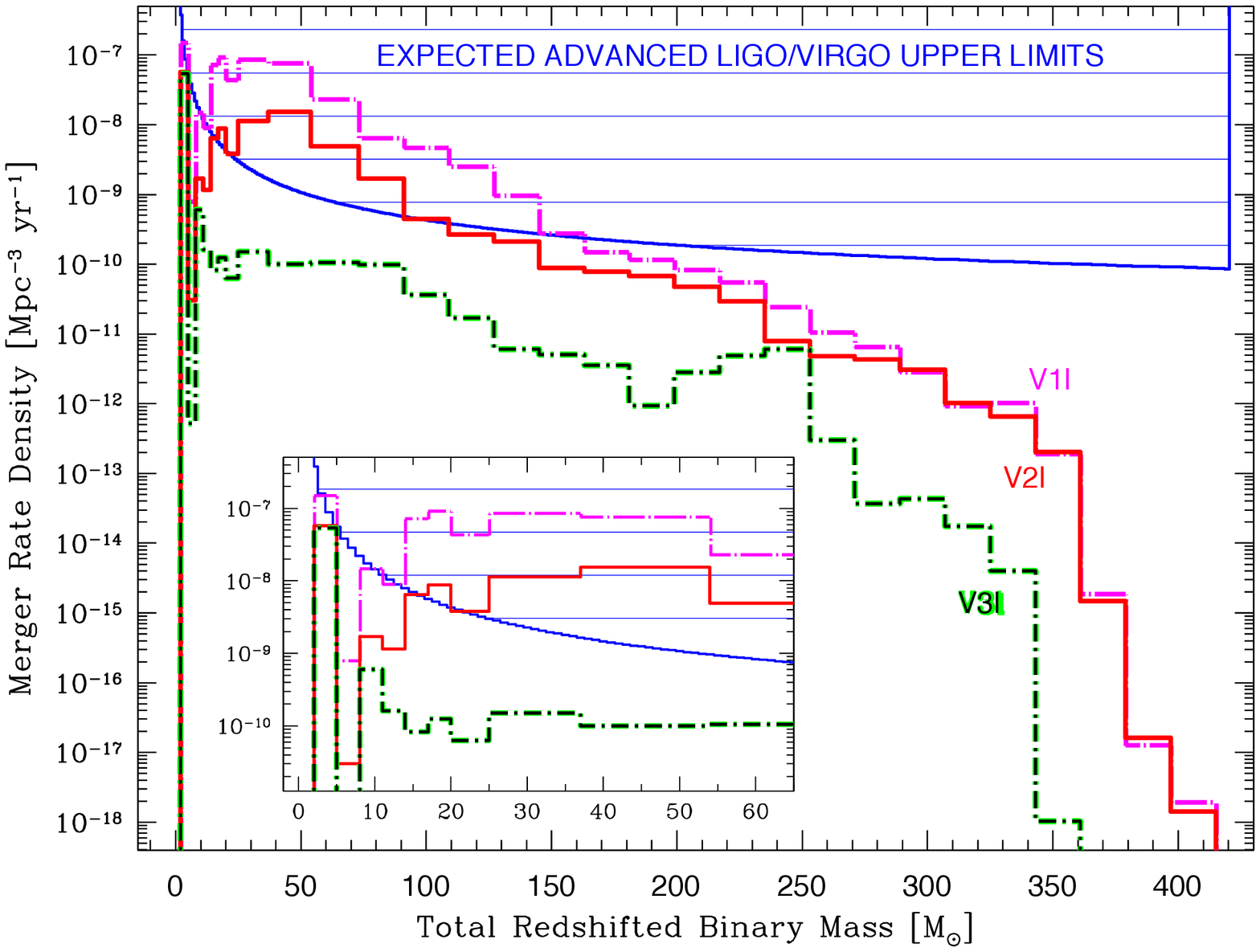}
\vspace*{-5.3cm}
\caption{
Merger rate density of double compact objects for our low metallicity
evolution scenario predicted (with method II) for advanced LIGO/Virgo. 
Note that the predictions for our optimistic (V1l) and standard (V2l) models  
are above the projected upper limits, while for our pessimistic model (V3l) 
the predictions are below the upper limits. The most likely detections are
predicted for BH-BH mergers with total redshifted mass in the range $25$--$73\msun$ 
(see the three highest mass bins, as compared to the upper limit curve).
}
\label{f4}
\end{figure}

\begin{figure}
\includegraphics[width=1.0\columnwidth]{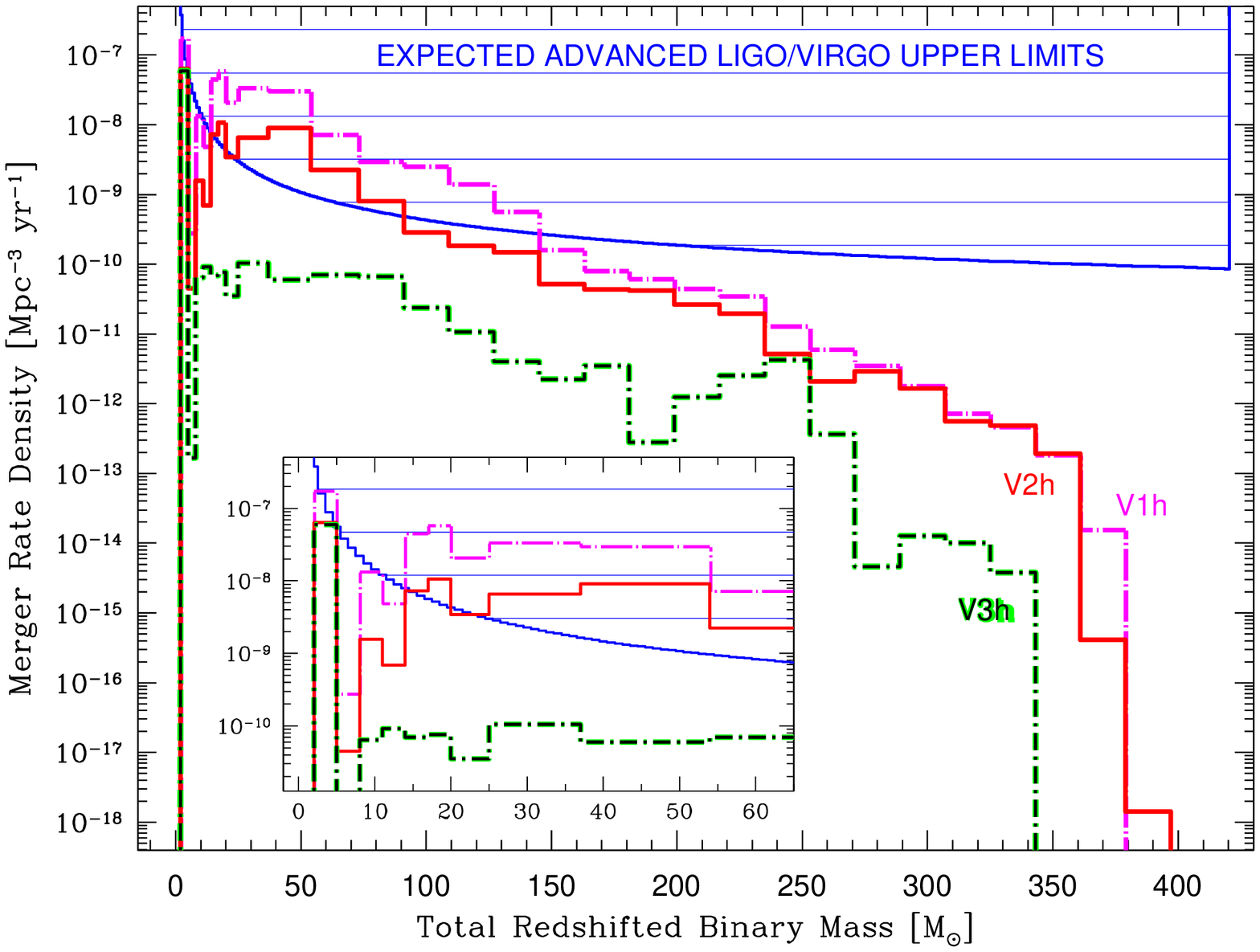}
\vspace*{-0.3cm}
\caption{
Merger rate density of double compact objects for our high metallicity
evolution scenario predicted (with method II) for advanced LIGO/Virgo. 
Note that the predictions for our optimistic (V1h) and standard (V2h) models  
are above the projected upper limits, while for our pessimistic model (V3h) 
the predictions are below the upper limits. The most likely detections are
predicted for BH-BH mergers with total redshifted mass in the range $25$--$73\msun$ 
(see the three highest mass bins, as compared to the upper limit curve).
}
\label{f5}
\end{figure}

Our conclusions and results are based on theoretical evolutionary calculations of Dominik 
et al. (2012, 2013, 2015) and are available online: \url{http://www.syntheticuniverse.org}. 
We find that the most likely sources to be detected with the advanced detectors are 
massive BH-BH mergers with total redshifted mass $\sim 30$--$70\msun$ (see Figs~\ref{f4} 
and ~\ref{f5} and Sec.~\ref{results2}). These massive mergers are predicted to consist of 
almost equal mass black holes; the mass ratio distribution peaks at $q=0.8$--$1.0$ (see
Fig. 8 of Dominik et al. (2015)). The specific evolution and physical properties of these 
massive BH-BH mergers are the subject of a forthcoming study (Belczynski et al., in prep.). 

This major finding is supported by the revised estimate for the stellar-origin maximum mass 
BH: $80\msun$ for stars below initial mass of $150\msun$ (Belczynski et al. 2010). Further, 
there is observational evidence that supports the existence of massive BH-BH binaries with 
short coalescence times. Bulik, Belczynski \& Prestwich (2011) argued that the two massive 
extragalactic binaries IC10 X-1 and  NGC300 X-1 are immediate progenitors of massive BH-BH 
mergers with intrinsic chirp mass in the range $\sim 10$--$30\msun$ (total intrinsic merger 
mass of $\sim 25$--$70\msun$ for equal mass black holes). The estimated empirical detection 
rate was so high that it was calculated that these BH-BH mergers could have been detected 
in LIGO/Virgo initial observations. 
The failure to detect  such binaries in S5/S6 LIGO/Virgo observations may be attributed 
to a potential overestimate of binary component masses in IC10 X-1 and NGC300 X-1. The mass 
determinations for these two binaries are subject to large uncertainty, as originally 
introduced and described by van Kerkwijk et al. (1996; see also our brief description of 
this issue in Sec.~\ref{ic10}). 

The above conclusion is derived from a majority of our models and is based on our estimates
and arguments for the best input stellar and binary physics. However, we have attempted to 
convey that the large uncertainties involved in evolutionary predictions do not allow us to 
make absolute predictions in favor of or against detections. We point out that the 
possibility remains that no BH-BH mergers will be detected even with the full sensitivity 
of advanced LIGO/Virgo (see below).

{\em (II) BH natal kicks and field/cluster massive BH-BH mergers.}
In Section~\ref{BHkicks} we argued that current electromagnetic constraints 
can not separate two basic trends: {\em (1)} natal kicks decrease with BH mass and {\em (2)} 
natal kicks are independent of BH mass. In Section~\ref{kicks} we argued that both basic natal 
kick mechanisms: {\em (i)} asymmetric mass ejection and {\em (ii)} asymmetric neutrino emission   
may have very similar signatures. For example, both of these mechanisms may generate natal kicks 
decreasing with BH mass, or both mechanisms may generate significant (in excess of $200\kms$) 
natal kicks for massive ($>10\msun$) BHs. This means that even if observations (whether 
electromagnetic or in gravitational radiation) will constrain the natal kick dependence on 
BH mass, the true physical mechanism generating the kicks may still remain a mystery. 
The only exception would be the case in which the natal kick increases with BH mass, as such 
the trend would be consistent {\em only} with the  asymmetric mass ejection mechanism in which 
asymmetry of ejecta grows with supernova delay time. Even if BH natal kick mechanism is not 
recognized, the measurement of the natal kick dependence on BH mass may provide very useful 
information for core collapse/supernova modeling. 

We have demonstrated in Section~\ref{results} that the two basic trends in the natal kick 
dependence on BH mass lead to very different predictions for BH-BH mergers. For natal kicks 
decreasing with BH mass, our standard evolutionary model predicts abundant detections of massive 
BH-BH mergers (with total redshifted mass $\sim 30$--$70\msun$) by advanced LIGO/Virgo. For natal 
kicks independent of BH mass our model with high kicks predicts that BH-BH merger detections are 
unlikely with advanced LIGO/Virgo. Finally, advanced LIGO/Virgo detections or non-detections 
of massive BH-BH mergers may constrain BH natal kicks, but they are  unlikely to provide 
information on the physical mechanism producing the natal kick.

Frequent detections of the massive BH-BH mergers will be indication that BHs receive low 
(if any) natal kicks. We have shown that high BH kicks severely limit LIGO/Virgo detections 
of BH-BH mergers originating from isolated (field) binary evolution (see Sec.~\ref{results}). 
In addition, although recent results show efficient BH-BH merger formation in globular clusters 
(Morscher et al. 20015; Rodriguez et al. 2015), the majority of these systems would similarly 
be disrupted if there are high BH kicks (M.~Morscher, 2015: Aspen Black Holes in Dense Star 
Clusters conference communication). 

In the case of low BH natal kicks (at least for massive BHs), abundant mergers are 
expected both from evolution of the field population ($\sim500$ detections per year; 
Dominik et al. (2015) standard input physics result) and from globular cluster 
evolution ($\sim100$ detections per year; Rodriguez et al. (2015) typical estimate).
Field massive BH-BH mergers are predicted to reach intrinsic (not redshifted) chirp 
mass of $50\msun$ (total intrinsic merger mass of $\sim 120\msun$ for equal mass 
black holes), have almost equal mass components and aligned spins. Our model with 
low BH natal kicks (natal kicks decreasing with BH mass) employs no mass ejection 
and no natal kicks for massive BHs, and therefore it naturally produces massive BH-BH 
mergers with aligned spins. This inference is only true if the two 
stars in the progenitor binary are formed with spins aligned with the binary angular 
momentum {\em and}\/ if no process (e.g., interactions with a third body) misaligns spins 
during binary evolution. 
Cluster massive BH-BH mergers are predicted to reach intrinsic chirp mass of $30\msun$ 
(total intrinsic merger mass of $\sim 70\msun$ for equal mass  black holes), have almost 
equal mass components (C.Rodriguez, private communication 2015) and misaligned spins. The 
misalignment is expected due to the dynamical captures/exchanges involved in the formation 
of BH-BH binaries and also due to the tilting of orbits during the dynamical encounters 
in dense star clusters. In the case of abundant detections, extracting spin information 
from the binaries may potentially provide a probe of the origin of the binary systems.

{\em (III) BH-BH non-detection case.} 
In the case of non-detection of massive BH-BH mergers by advanced LIGO/Virgo, it may not 
be clear  which physical process is most responsible for limiting BH-BH binary formation.  
Two suppression mechanisms have been identified and discussed (see Sec.~\ref{nobhbh}): 
high BH natal kicks (this study) and strong stellar winds (Mennekens \& Vanbeveren 2014). 
However, it is worth noting that strong winds should not affect dynamically formed BH-BH 
mergers. Therefore detections with confirmed cluster origin combined with non-detection 
from field populations would in principle clarify this issue. Of course, other inhibitory 
processes may also exist.

{\em (IV) Very massive stellar origin BH-BH mergers.} 
The above considerations apply only to the evolution of stars with initial mass below 
$150\msun$. Recent observations of the R136 cluster in the LMC ($\sim 0.6 \zsun$)
identified stars with masses larger than $150\msun$ (Crowther et al. 2010). This may be an 
indication that the IMF extends to much higher masses that were considered previously 
{\em only} for primordial (Population III) stars in the high redshift Universe. If this is 
the case, there may exist an additional population of very massive stellar origin BHs with 
mass $\gtrsim 100\msun$ (Yusof et al. 2013). Formation of close BH-BH systems from these 
massive stars may be possible, although estimates are subject to severe uncertainties. If 
detections of such objects are made, they will provide information on the extent of the IMF 
and the occurrence of pair instability supernovae (Belczynski et al. 2014).

\acknowledgements
We would like to thank a number of colleagues who have helped us to improve our project
over the past several years. We want to name not only those who have given us information 
and positive feedback, but also those who have provided critiques.
The full list includes Ilya Mandel, Gijs Nelemans, Selma de Mink, Felix Mirabel, 
Tsvi Piran, Patric Brady, Alessandra Buonanno, David Reitze, Vicky Kalogera, Tassos 
Fragos, Megan Morscher, Carl Rodriguez, Scott Hughes, Tom Maccarone, Yizhong Fan, and 
Dany Vanbeveren.
KB and MD acknowledge support from the NCN grant Sonata Bis 2 (DEC-2012/07/E/ST9/01360)
and FNP professorial subsidy MASTER 2013. TB and KB acknowledge support from the NCN 
grant Harmonia 6 (UMO-2014/14/M/ST9/00707). This work was supported in part by National 
Science Foundation Grant No. PHYS-1066293 and the hospitality of the Aspen Center for 
Physics (KB, DH). This work was partially supported by a grant from the Simons 
Foundation (KB).
ROS was supported by \textsc{nsf} grant PHY-1505629.
DEH was supported by \textsc{nsf career} grant PHY-1151836. He also acknowledges support
from the Kavli Institute for Cosmological Physics at the University of Chicago through 
\textsc{nsf} grant PHY-1125897 as well as an endowment from the Kavli Foundation.
EB is supported by \textsc{nsf career} grant PHY-1055103 and by FCT Contract 
IF/00797/2014/CP1214/CT0012 under the IF2014 Programme.
The work of CF was done under the auspices of the National Nuclear Security 
Administration of the U.S. Department of Energy, and supported by its contract 
DEAC52-06NA25396 at Los Alamos National Laboratory.
TB was supported by the NCN grant UMO-2014/15/Z/ST9/00038.

\clearpage

\begin{deluxetable}{ccl}
\tablewidth{350pt}
\tablecaption{Evolutionary Models\tablenotemark{a}}
\tablehead{Model & Name & Description} 
\startdata
V1 & optimistic    & in CE: HG donors allowed \\
V2 & standard      & $\lambda=$physical, decreased BH kicks, \\
   &               & rapid SN, HG CE donors not allowed\\ 
V3 & pessimistic   & full natal kicks for BHs \\
V4 & no mass-gap   & delayed supernova engine \\
\enddata
\label{t2}
\tablenotetext{a}{
For each model (V1--V4) we provide merger rate densities in low-metallicity (l: e.g.,
V1l) and high-metallicity (h: e.g., V2h) evolution scenarios.}
\end{deluxetable}

\begin{deluxetable}{llll}
\tablewidth{350pt}
\tablecaption{Black Hole Mass and Natal Kick \tablenotemark{a}}
\tablehead{No & Name & Mass [$\msun$] & Natal Kick [km s$^{-1}$]}
\startdata
1)  & GRO J1655-40 (Nova Sco 94) & $5.3 \pm 0.3$   (3,21) & $0$--$210$ (4) \\
2)  & XTE J1118+480              & $7.6 \pm 0.7$    (10)  & $80$--$310$ (11) \\
3)  & GS 2023+338 (V404 Cyg)     & $9.0_{-0.6}^{+0.2}$ (17) & $0$--$45$ (23,1)\tablenotemark{b} \\
4)  & GRS 1915+105               & $12.4_{-1.8}^{+2.0}$ (13)\tablenotemark{c} & $0$--$75$ (24,13,1)\\
5)  & Cyg X-1                    & $14.8 \pm 1.0$   (18)  & $0$--$60$ (19,20) \\  
&&&\\    
6)  & 4U 1543-47                 & $5.1 \pm 2.4$    (22)  & $>80$ (2), $>75$ (1) \\ 
7)  & H 1705-250 (Nova Oph 77)   & $6.4 \pm 1.5$     (5)  & $>217$ (6)\tablenotemark{d}, $>0$ (1)\\ 
8)  & A 0620-00 (V616 Mon)       & $6.6 \pm 0.25$    (8)  & $>20$ (6), $>0$ (1) \\
9)  & GRS 1124-68 (Nova Mus 91)  & $7.0 \pm 0.6$     (9)  & $>62$ (6), $>0$ (1)\\
10) & GS 2000+251                & $7.2 \pm 1.7$     (7)  & $>24$ (6), $>0$ (1)  \\
11) & GRS 1009-45 (Nova Vel 93)  & $8.5 \pm 1.0$    (12)\tablenotemark{e} & $>49$ (6), $>0$ (1)\\
12) & XTE J1819-254 (V4641 Sgr)  & $10.2 \pm 1.5$   (14)  & $>190$ (2), $>100$ (1) \\
13) & GRO J0422+32               & $>10.4$          (15)  & $>35$ (6), $>0$ (1)\\
14) & XTE J1550-564              & $10.5 \pm 1.0$   (16)  & $>10$ (2), $>0$ (1) \\

\enddata
\label{T:kicks}
\tablenotetext{a}{
References for the mass and natal kick estimates are given in parentheses:
(1) this study; see Sec.~\ref{BHkicks},
(22) Orosz et al. (1998), (2) Repetto et al. 2012, (3) Motta et al. (2014),   
(4) Willems et al. (2005), (5) Harlaftis et al. (1997), (6) Repetto \&
Nelemans (2015), (7) Ioannou et al. (2004), (8) Cantrell et al. (2010), 
(9) Gelino et al. (2001), (10) Khargharia et al. (2013), (11) Fragos et al. 
(2009), (12) Macias et al. (2011), (13) Reid et al. (2014), (14) Orosz  
et al. (2001), (15) Reynolds et al. (2007), (16) Li et al. (2013), (17) 
Khargharia et al. (2010), (18) Orosz et al. (2011), (19) Mirabel \& Rodrigues (2003), 
(20) Wong et al. (2012), (21) Beer \& Podsiadlowski 2002, (23) Miller-Jones et al.
2009, (24) Dhawan et al. (2007).}
\tablenotetext{b}{The quoted range refers to our own simple estimate based on 3D 
peculiar velocity presented in (23).}
\tablenotetext{c}{There is also an alternative BH mass estimate of $12.9\pm2.4\msun$
(Hurley et al. 2013). These observations were superseded with much higher resolution
spectra and analysis of Steeghs et al. (2013) that resulted in a BH mass of $10.1\pm0.6\msun$. 
Recently revised system parallax/distance led to further correction of the BH
mass to the value included in this Table.}
\tablenotetext{d}{Reported lower limit of $415\kms$ (6) was revised later by the first 
author of (6) to $217\kms$. The revision is a result of a mistake made in calculation 
for just this one system in (6).}
\tablenotetext{e}{Error estimate: Jerome Orosz 2014, priv. communication.}
\end{deluxetable}

\pagebreak

\pagebreak
\begin{figure}
\vspace*{-3cm}
\includegraphics[width=1.0\columnwidth]{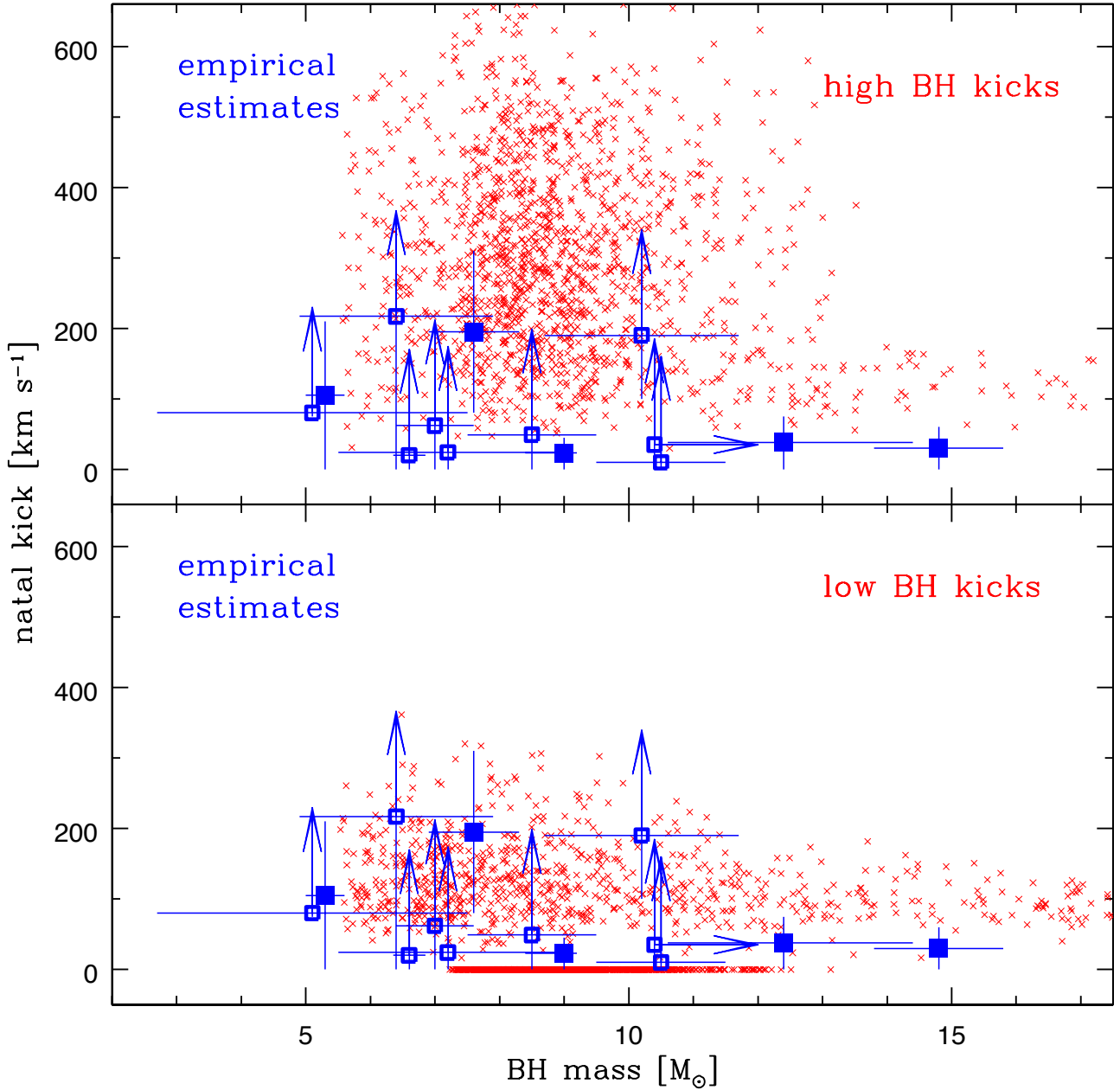}
\vspace*{-4.5cm}
\caption{
Black hole natal kick empirical estimates (blue; squares) for 14 Galactic X-ray 
binaries. Actual natal kick estimates based on 3D peculiar velocities are available for 5 
binaries (filled squares). Lower limits based on Galactic position are available for 9 
binaries (empty squares with up-arrows). We have re-evaluated some of these lower limits down 
(lines reaching from the original lower limit down to the revised limit). The most extreme 
case is H 1705-250 with an original lower limit of $217\kms$ and a revised limit of $0\kms$. 
Errors on BH mass estimates are marked with horizontal lines. For one system (GRO J0422+32) 
there is only a lower limit on the BH mass ($>10.4\msun$), marked with a right-arrow. The 
empirical data used in this plot is given in Table~\ref{T:kicks} and described in
Sec.~\ref{BHkicks}. 
The empirical estimates are contrasted with evolutionary predictions (red; small crosses) of 
Galactic BH interacting binaries. In one evolutionary model we have assumed high BH kicks 
(top panel): all BHs receive natal kicks as measured for single Galactic pulsars (Maxwellian 
with 1D $\sigma=265\kms$: an average kick of $420\kms$). In another model (bottom panel) we 
have adopted low BH kicks (approximately decreasing with BH mass; rapid supernova explosion 
model of Fryer et al. 2012). 
Contrary to some expectations that natal kicks decrease with BH mass, we point out that 
both theoretical models: natal kicks independent of BH mass and kicks decreasing with BH mass, 
can explain the empirical data within their associated errors. On one hand this reflects the 
fact that the empirical data is still very poor (only 5 good estimates and many weak lower 
limits). On the other hand this demonstrates that even model with natal kicks decreasing with 
BH mass may deliver a wide range of BH natal kicks. For low-mass BHs low and high kicks are 
expected as we draw them from Maxwellian with high $\sigma\approx265\kms$. For high mass BHs, 
low to zero kicks are expected if a BH forms with high mass ($\sigma\approx0\kms$), and higher 
kicks are expected for a BH that has formed at low-to-intermediate mass 
($\sigma\approx0$--$260\kms$) and then increased its mass via accretion from its companion.   
} 
\label{bhkick}
\end{figure}

\clearpage

\appendix

\section{A. Black hole natal kick estimates}
\label{Appkicks}

\subsection{Natal Kicks From Galactic Positions: I}
\label{RN2015}

Repetto \& Nelemans (2015) have studied natal kicks in seven Galactic BH low mass X-ray 
binaries (BH LMXBs) with very short periods ($<1$d; thus only with main sequence donors). 
These include: GRO J0422+32, GRS 1009-45, A 0620-00, GS 2000+251, GRS 1124-68 (Nova Mus 91 
in their paper), XTE J1118+480, H 1705-250. The Galactic plane ($h_{\rm z}=0$~kpc) was assumed 
to be the birth place of these binaries. Positions of these binaries above the Galactic plane 
($h_{\rm z}$) along with a simple model of binary physics were used to estimate lower 
limits on natal kicks. The results indicate rather low natal kick lower limits for five 
systems ($\lesssim 60\kms$), while for one system (XTE J1118+480) the lower limit is 
somewhat higher ($\sim 100\kms$) and for another one (H 1705-250) the lower limit is very 
high ($\sim 400\kms$). It was concluded that BHs can receive both low and high natal 
kicks and that natal kicks do not correlate with BH mass within this small subpopulation 
of BH systems. The results were found for binaries reported to host BHs with mass in the 
range: $3.0$--$8.8\msun$. 

The BH masses may be potentially revised to a higher range: from $5$ to above $10\msun$ 
(see our Table~\ref{T:kicks}). 
The BH mass of GRO J0422+32 was revised from $3$--$5\msun$ (Gelino \& Harrison 2003) to a 
mass larger than $10\msun$ (Reynolds et al. 2007). Both estimates are based on different 
information to establish inclination (and thus BH mass) of this binary. The former estimate 
is based on the best observations of ellipsoidal modulations, while the latter employs  
the best estimate of the disk contribution (found at $\sim 30\%$ level) to the light 
curve. The BH mass of GRS 1009-45 estimated at first at $4.4\msun$ (Filippenko et al. 1999),  
was later revised with new spectroscopic and photometric observations to $8.5\msun$ (Macias 
et al. 2011). This revised range of BH masses does not require BHs within the mass gap, and it 
includes both Galactic low mass BHs ($5$--$10\msun$) and high mass BHs ($10$--$15\msun$).  

H 1705-250, depending on the large distance estimate uncertainty ($6.5$--$10.7$~kpc), is 
located at $h_{\rm z}=1.0$--$1.7$~kpc above the Galactic plane right in the bulge region. Since 
the bulge scale height of stars is about $0.7$~kpc (e.g., Juric et al. 2008; SDSS data), 
there is a non-negligible probability that H1705-250 was born at (or nearby) its current 
location. The donor star in this system is a low mass main sequence star 
($\sim 0.15$--$1.0\msun$; Martin et al. 1995, Filippenko et al. 1997), thus it cannot be 
excluded that this system was born longer than $10$~Gyr ago in the bulge, with no natal kick 
needed to place the binary where it is found today.
On the other hand, it can not be excluded that this system was born in the thin disk as 
assumed by Repetto \& Nelemans (2015). The thin disk current scale height is $\sim 0.3$~kpc 
(Juric et al. 2008), and the most likely scale height for the formation of X-ray binaries 
is $\sim 0.1$kpc (e.g., Brandt \& Podsiadlowski 1995). The calculation of Repetto \& Nelemans 
(2015) with the assumed initial position of binary at the Galactic plane ($h_{\rm z}=0$~kpc) 
resulted in a minimum natal kick of $415\kms$ for the BH in H 1705-250. However, there was a 
mistake in this estimate (S.Repetto, private communication 2015) and the corrected lower 
limit is now set at $217\kms$. We keep this estimate and mark it with a point and upward arrow 
in Figure~\ref{bhkick}, to indicate a lower limit that corresponds to the initial location of 
this system at the Galactic plane. But we also allow for the possibility that this system was 
born in the bulge without a natal kick: we mark our potential revision of the Repetto \& 
Nelemans (2015) estimate with a line going all the way down to $0\kms$.

The other five systems (GRO J0422+32, GRS 1009-45, A 0620-00, GS 2000+251, GRS 1124-68) have 
small estimated distances from the Galactic plane: $h_{\rm z}<0.8$~kpc (Repetto \& Nelemans 
2015). All of these systems host currently low mass donors (typically $\sim 0.5\msun$ stars; 
\url{http://www.stellarcollapse.org/bhmasses}), so 
it can not be excluded that they were born longer than $11$ Gyr ago in the thick disk.   
Since the thick disk current scale height is $1$~kpc (e.g., Juric et al. 2008) all these 
systems could have been potentially born at (or near) their current location. This would 
imply no natal kick needed for BHs in these binaries. However, the donor stars in these 
binaries may have started with much higher mass ($\gtrsim 1 \msun$) making them younger than 
the thick disk population, or the thick disk was initially much more compact (in vertical scale)
than today. In such cases the estimates of Repetto \& Nelemans (2015), that place these 
binaries in the Galactic plane, are correct. As in the case of H 1705-250 we mark both 
possibilities  in our Figure~\ref{bhkick} and Table~\ref{T:kicks}.    

The only exception (in the Repetto \& Nelemans 2015 sample) from this systematic uncertainty 
on the initial source position is XTE J1118+480 located at a distance of $1.6$--$1.8$~kpc and 
at $h_{\rm z}=1.4$--$1.6$~kpc above the Galactic plane (so it is not in the bulge nor the 
thick disk). However, the estimate based only on its position is superseded by one that 
includes its known 3D velocity (see Sec.~\ref{xtej1118}).

\subsection{Natal Kicks From Galactic Positions: II}
\label{R2012}

There are  four additional BH binaries with natal kick lower limits derived from
their positions in the Galaxy studied by Repetto, Davies and Sigurdsson (2012). 
In this work, Repetto et al. (2012) computed the minimum natal kick needed in 
order to place the binary at the current height from the Galactic plane. They 
assume the binary to be born in the Galactic plane (at $h_{\rm z}=0$~kpc), and they 
take the central value of the distance to the BH systems to calculate their 
current positions above the Galactic plane. Unlike the work by Repetto et al. 
(2015), they do not follow the binary evolution of the BH systems, but rather
they take average binary properties.

Here we revise their estimates accounting for the scale height of the disk, and 
allowing for uncertainty of the distance estimate. Note that in a different 
part of their study, Repetto et al. (2012) have carried a binary population 
synthesis calculation and allowed for a spread of initial positions of 
progenitor binaries in the thin disk with a scale height of $0.167$~kpc. They 
followed motion of binaries in the Galactic potential to check what natal kick 
distribution is needed to recover overall $h_{\rm z}$-distribution of the BH binaries. 
This part of their study was not used to estimate natal kick lower limits
for particular binaries and we do not use it. 

In the case of XTE J1550-564 ($h_{\rm z}=-0.12$ to $-0.15$~kpc, $M_{\rm opt}=0.3\msun$
is the current mass of a donor star), this binary has a small distance from the Galactic 
plane, which is compatible with the formation within the scale height $h$ of the disk 
where X-ray binaries are formed ($h \sim 0.1$~kpc; Brandt \& Podsiadlowski 1995), and 
which does not require a natal kick at birth. We therefore revise  the lower limit of 
$10\kms$ calculated by Repetto et al. (2012), pushing it down to $0$ km/s.

In the case of GS 2023+338, the 3D peculiar velocity was measured and a better estimate 
of the natal kick may now be obtained, superseding the value presented in Repetto et al. 
(2012). See the discussion in Sec.~\ref{gs2023}.  

In the case of XTE J1819-254 ($h_{\rm z}=-0.63$ to $-1.03$~kpc; the spread in $h_{\rm z}$ 
reflects uncertainty on distance estimate) its average height from the Galactic plane 
($h_{\rm z}=-0.83$~kpc) lead to a natal kick estimate of $190\kms$ in Repetto et al. (2012). 
The companion star to the BH has a mass $M_{\rm opt}=5.5$--$8.1\msun$ and a spectral type
and luminosity class: B9III (\url{http://www.stellarcollapse.org/bhmasses}). 
Its age being young ($<100$ Myr), one can exclude its origin in 
thick disk or bulge. We therefore place the progenitor at $h_{\rm z}=0.3$~kpc that 
corresponds to the current scale height of the thin disk (Juric et al. 2008). We then follow 
the same analysis as Repetto and Nelemans (2015) to estimate the minimum binary center-of-mass 
velocity to move it to the minimum height estimated for the current position of this system at 
$h_{\rm z}=-0.63$. This velocity is assumed to be entirely in $-h_{\rm z}$ direction and 
the motion is tracked in Galactic potential. Our estimate: $100\kms$ provides a conservative 
lower limit on BH natal kick. The natal kick has to be larger than the binary center-of-mass 
velocity, as natal kick is received by a $\sim 10\msun$ BH while the center-of-mass 
velocity operates on entire binary with mass of $\gtrsim 15\msun$. 

In the case of 4U 1543-47 ($h_{\rm z}=0.70$~kpc, $M_{\rm opt}=2.3$--$2.6\msun$; A2V), with the 
optical component/donor star maximum age of $\sim 1$ Gyr, the thick disk or bulge origin 
may be eliminated. Repetto et al. (2012) have set a lower limit on natal kick ($80\kms$) 
assuming the binary is formed right at the Galactic plane ($h_{\rm z}=0$~kpc). Using the 
same arguments as in the case of XTE J1819-254, we reset this lower limit to $75\kms$.  

For XTE J1550-564, XTE J1819-254 and 4U 1543-47, we keep original Repetto et al. (2012) 
estimates, but we mark our modest revisions both in Table~\ref{T:kicks} and in 
Figure~\ref{bhkick}.

\subsection{Natal Kick From 3D Velocity: GRO J1655-40}
\label{gro}

Willems et al. (2005) have performed analysis of available evolutionary, nucleosynthetic and 
proper motion constraints for GRO J1655-40. They have backtracked 3D motion and evolution of 
this binary to the time of BH formation in the Galactic potential to estimate post supernova 
velocity. This post supernova velocity was in turn used to estimate the magnitude of a BH 
natal kick. The estimated 3D post supernova velocity ($45$--$115\kms$) may potentially lead 
to much smaller natal kick velocity, than earlier estimates based just on present-day radial 
velocity ($-114\kms$; corrected for the Sun motion and Galactic differential rotation).    

Evolutionary predictions along with Galactic motion constraints yield a wide range of BH 
natal kick values: $0$--$210\kms$. Willems et al. (2005) quote this range for BH mass 
($5.4\msun$) adopted from Beer \& Podsiadlowski (2003). We chose this model as this mass was 
later confirmed by X-ray timing analysis (Motta et al. 2014; $5.31\pm0.07\msun$). Note that 
symmetric supernova scenario ($0\kms$ natal kick) is allowed in this solution and that the 
present-day 3D velocity is fully accounted for by the Blaauw kick (Blaauw 1961; extra 
systemic velocity just from mass loss during BH formation) and Galactic orbit of GRO J1655-40. 
Also, note that the largest allowed BH natal kick is just half of what is estimated for 
Galactic pulsars ($420\kms$; average NS 3D natal kick from Hobbs et al. 2005).   

Additional, supernova/nucleosynthetic constraints led Willems et al. (2005) to impose a 
somewhat more stringent constraint on the BH natal kick: $40$--$140\kms$. This conclusion 
was based on detection of enhanced abundance of some heavy elements (O, Mg, S, Si, Ti; $10$ 
times solar) in the GRO J1644-40 low-mass donor star spectrum. These overabundances were 
used to argue for supernova (with its mass loss assumed to have polluted companion star) to 
form the BH in GRO J1644-40 (Israelian et al. 1999). Willems et al. (2005) have used 1D 
supernova models (Fryer et al. 1999) to estimate that only helium stars in narrow mass 
window $8$--$10\msun$ may both produce $\sim 5\msun$ BH with enough asymmetry to deposit the 
heavy elements in the GRO J1644-40 optical star. The limits on helium star mass enabled to 
narrow down the ``successful'' evolutionary sequences (and natal kicks) leading to the 
formation of binaries resembling GRO J1644-40. 

Supernova models with the associated nucleosynthetic yields employed by Willems et al. (2005) 
in fact produce titanium farther out from the  exploding star center for the  massive helium 
star model ($10\msun$) as compared with the lower mass model ($6.2\msun$). Thus they allow for 
easier deposition of this element in the companion atmosphere (see their Fig.10). However, 
in both models, iron production shows very similar trend as titanium does. If we are to 
believe the models and arguments presented by Willems et al. (2005) there should be an 
overabundance of iron on par with the observed overabundance of titanium. However, the same 
observations used to measure the overabundance of titanium resulted with approximately 
solar iron abundance in GRO J1644-40 ($[Fe/H]=0.1\pm0.2$; Israelian et al. 1999). 

Willems et al. (2005) ``successful'' evolutionary sequences are based on details of RLOF 
calculations and response of the donor to mass loss. Some of their sequences show that RLOF 
could be initiated in eccentric binaries (e.g., their Fig.~5), but they are circularized and 
synchronized instantaneously (``by hand'') at the onset of RLOF. However, it is not clear how 
to treat such cases. It seems that for some binary configurations RLOF may rather quickly 
circularize the orbit, while for others the eccentric periodic mass transfer at periastron 
may be a rather prolonged phase (Sepinsky et al. 2010).  

The mass loss from the optical star in GRO J1644-40 was quite significant in models presented 
by Willems et al. (2005).  The evolutionary sequences allowed for a change of initial star 
mass from $2.3$--$4.0\msun$ down to $1.45\msun$. Such a significant mass loss and its effects 
on star evolution (apparent aging, decrease in burning rates, radius evolution and its visual 
properties) are not fully understood. This just reflects the fact, that the stellar models are 
still subject to large uncertainties arising from treatment of convection, mixing and rotation. 
For example, some particular stellar models were adopted for Willems et al. (2005) estimates 
(e.g., models without rotation and with moderate convection; Ivanova et al. 2003), but the 
systematic uncertainties involved in stellar evolution were not tested. 

The significant mass loss predicted by Willems et al. (2005) models and the fact that donor 
star is massive enough to have radiative envelope (not as much mixing as in case of convective 
envelope) it is surprising that observed overabundance of elements is considered to be a 
deposition from supernova explosion. The significant RLOF mass loss should have removed the 
polluted outer layers from the donor star. It seems that either the donor star has not lost 
significant mass in RLOF (and the models are incorrect) or that the observed overabundances 
are not connected to supernova explosion that has formed the BH in this system. 

Despite of the evolutionary uncertainties Willems et al. (2005) is at the moment the most 
thorough and detailed study of BH natal kick in GRO J1644-40. Due to the caveats discussed 
above we are hesitant to use the more stringent constraints on BH natal kick. We chose to 
adopt the wider range presented by Willems et al. (2005). It is very informative, and possibly 
contrary to some preconceptions in the community, that the BH in this system is allowed to 
have formed {\em without} any natal kick ($0\kms$). At best, this BH has formed with a 
{\em moderate} natal kick ($210\kms$).

\subsection{Natal Kick From 3D Velocity: XTE J1118+480}
\label{xtej1118}

Fragos et al. (2009) have studied available constraints, 3D peculiar velocity included, on 
the past evolution of XTE J1118+480. Using methods very similar to the ones employed for 
GRO J1655-40 (Willems et al. 2005) the past history and the BH natal kick were reconstructed.  
The major difference between the two studies comes from the fact that nucleosynthetic 
information was not used to constraint the natal kick by Fragos et al. (2009) as the
abundance measurements for XTE J1118+480 do not show such striking  features as in the 
case of GRO J1655-40. This actually makes this case stronger (see Sec.~\ref{gro} for 
discussion of issues with nucleosynthetic constraints).

Fragos et al. (2009) estimated that current properties of the system, with its 3D 
peculiar velocity, combined with RLOF sequences and evolutionary predictions for low 
mass MS donors, constrain the BH natal kick in XTE J1118+480 to: $80$--$310\kms$.
This estimate has two potential caveats, both discussed by authors. 

The wider set of donor stars could be potential progenitors of donor stars than 
considered by Fragos et al. (2009; only MS stars with mass $<1.6\msun$). 
Intermediate mass donors ($\sim 3$--$5\msun$) were also proposed to be potential 
progenitors of BH low-mass X-ray binaries (e.g., Justham, Rappaport \& Podsiadlowski 2006).
However, models of such massive donors (stripped from most of their mass by RLOF) seem 
not to match effective temperatures of observed stars, while Fragos et al. (2009) can 
match the temperature of donor in case of XTE J1118+480. If somehow BH LMXBs originated 
from intermediate mass donors, that would lead to different evolutionary constraints 
on the formation of XTE J1118+480 and thus affect constraints on BH natal kick.  

Fragos et al. (2009) assumed that the system originated in the Galactic thin disk and that 
the donor star had solar metallicity. This assumption, coupled with significant 3D peculiar 
velocity and its current position in halo, led directly to the estimate of high natal kick 
for XTE J1118+480. The alternative scenario, that does not require significant natal BH kick 
(if any natal kick at all), allows for the formation of this system in halo. Although this 
scenario can not be excluded, it does not seem very likely. Only two Galactic globular 
clusters have solar metallicity: Terzan 5 and Liller 2 (e.g., Harris 1996). More 
convincingly, Gonzalez Hernandez et al. (2006) detected super-solar abundances in the donor 
atmosphere. In a follow-up study, Gonzalez Hernandez et al. (2008) explored a series of SN 
explosion models in order to match the observed abundances of the donor atmosphere. They 
found the best match for a donor born in the Galactic thin disk, hence with solar metallicity, 
and later polluted with the ejecta of the exploding companion. A birth in the halo provided 
instead unacceptable fits to the observed abundances.

\subsection{Natal Kick From 3D Velocity: GS 2023+338 (V404 Cyg)}
\label{gs2023}

GS 2023+338 is the system qualitatively resembling GRS 1915+105 (see Sec.~\ref{grs1915}). 
It consists of a $9\msun$ BH with a low mass and evolved companion 
($M_{\rm opt} \sim 0.5$--$1.0\msun$; K0/3 IV \url{http://www.stellarcollapse.org/bhmasses}) 
on a rather large orbit ($P_{\rm orb}=6.5$d). Miller-Jones et al. (2009a) presented parallax 
determination, that led to a revised distance to the source ($2.39\pm0.14$~kpc). Using this 
precise distance estimate they have used radial velocity and proper motion known for this 
system to derive 3D peculiar velocity of $39.9\pm5.5\kms$. Miller-Jones et al. (2009a) 
concluded that the Galactic plane component of this velocity: $39.6\kms$ is well above 
dispersion velocity in the position of GS 2023+338 ($18.9\kms$). This implies that the 
GS 2023+338 3D peculiar velocity (or some significant part of it) originates from the BH 
formation. They also argued that Blaauw kick is most likely responsible for this 3D peculiar 
velocity, with natal kick (if any) being small. 

We disagree with the above conclusion. In fact, it is allowed that the Blaauw kick (just 
symmetric mass loss) from the system produces peculiar velocity of $\sim40\kms$ and that 
would require mass loss at supernova at the level of $5$--$10\msun$ (e.g., Miller-Jones et 
al. 2009b). However, it is also allowed that this $9\msun$ BH has formed without any 
significant mass loss as BH formation mechanism is not yet fully understood. If this was the 
case than natal kick could have provided most of the peculiar velocity (with some correction 
for potential contribution from dispersion velocity). Therefore, we adopt a range of 
$0$--$45\kms$ for the value of natal kick in GS 2023+338; from no natal kick to the maximum 
estimate on peculiar velocity from Miller-Jones et al. (2009a).

\subsection{Natal Kick From 3D Velocity: GRS 1915+105}
\label{grs1915}

Dhawan et al. (2007) have obtained the estimate of 3D peculiar velocity from parallax and 
proper motion measurements for GRS 1915+105. For the revised distance to the source 
($7$--$10.6$~kpc; Reid et al. 2014) the 3D peculiar velocity was estimated at the level of 
$\sim 45$--$75\kms$ (see Fig.3 of Dhawan et al. 2007). Such a small velocity was interpreted 
as a result of Galactic velocity diffusion implying BH natal kick of $0\kms$. 
Another measurement of GRS 1915+105 proper motion and parallax translated to a peculiar 
3D (non-circular/corrected for Galactic rotation) velocity of $22\pm24\kms$ and a $95\%$ 
confidence upper limit of $61\kms$ (Reid et al. 2014). 

Since there is no comprehensive study of GRS 1915+105 evolution and motion in Galaxy we adopt 
the upper limit on peculiar 3D velocity as an upper limit on natal kick velocity (no diffusion, 
and no mass loss at BH formation assumed) and we allow for the no-natal-kick possibility (the 
diffusion scenario, or entire peculiar velocity from Blaauw kick assumed). The range for natal 
kick in GRS 1915+105 is then: $0$--$75\kms$. This is only an approximation as natal kick may 
have been higher than the upper value of peculiar velocity in case the Blaauw kick and natal 
kick worked in opposite directions. 

Systemic peculiar velocity changes (in quasi-periodic way) in time as binary moves on its 
complex orbit in Galactic potential. For example, in the case of XTE J1118+480 the peculiar 3D 
velocity changes from $\sim 100\kms$ to $\sim 200\kms$ during motion of this binary in Galaxy 
(Fragos et al. 2009). GRS 1915+105 is on more circular and at lower $h_{\rm z}$ orbit (e.g., 
Dhawan et al. 2007) than XTE J1118+480 so the changes are not expected to be as drastic. 
However, the GRS 1915+105  peculiar velocity may have been much higher right after the BH 
formation and then it was gradually ``thermalized'' by interactions with much slower stars. 
After all, GRS 1915+105 moves in the relatively dense Galactic disk and it was very likely 
formed many Gyr ago (e.g., Belczynski \& Bulik 2002). This just demonstrates the dangers of 
using the present day 3D peculiar velocity as a proxy for natal kick (as we do in the case of 
GS 2023+338 or GRS 1915+105) or as a starting point for BH natal kick estimates (e.g., 
Willems et al. 2005 or Fragos et al. 2009).

\subsection{Natal Kick From 3D Velocity: Cyg X-1}

Wong et al. (2012) have studied evolution and dynamics of 3D motion of Cyg X-1
based on parallax measurement of Reid et al. (2011).
They have used the same method as Willems et al. (2005) and Fragos et al. (2009)
in study of GRO J1655-40 and XTE J1118+480, respectively. 
Using position of this system in Galaxy, its radial velocity and its proper motion 
Wong et al. (2012) traced this system back in time to the moment of BH formation. 
The system was found to be $h_{\rm z}=30$--$110$ pc above Galactic plane and move 
with peculiar 3D velocity of $22$--$32\kms$ right after BH formation. Allowing for 
mass ejection and natal kick at the BH formation, it was found that BH in Cyg X-1 may 
have formed with or without natal kick. The allowed range for natal kick is rather 
broad $0-115\kms$, with a peak probability of $20\kms$.

Mirabel \& Rodrigues (2003) argued that Cyg X-1 was born in Cyg OB3 association of 
massive stars due to their proximity and very similar motion on the sky. The space 
velocity of Cyg X-1 in respect to the association was estimated at $9\pm2\kms$. Wong 
et al. (2012) redid their analysis under assumption that the peculiar velocity of Cyg 
X-1 right after the BH formation is $\leq10\kms$. The allowed range for natal kick is 
found in the range $0$--$60\kms$. Probability distribution of natal kicks is not 
given in this case by Wong et al. (2012).

It can not be excluded that Cyg X-1 has originated from some other place than Cyg OB3 
association. However, the circumstantial evidence; similar distance, similar motion and the 
fact that massive stars are typically born in associations seems to be very convincing. 
Therefore, we adopt a natal kick range of $0$--$60\kms$ that corresponds to the origin 
of Cyg X-1 in Cyg OB3 association.

\subsection{Natal kicks for extragalactic BHs}
\label{ic10}

Note that there is one natal kick estimate ($<130\kms$; Wong et al. 2014) for an extragalactic 
X-ray binary IC10 X-1 with a potentially very massive BH ($\sim 20-30 \msun$; Prestwich et 
al. 2007; Silverman \& Filippenko 2008). Since this system does not enter our comparative 
analysis presented in the next section (Sec.~\ref{compkicks}), we do not discuss this case in 
detail. 

BH mass estimate for this system is based on semi-amplitude of HeII 4685.8A emission line 
(e.g., Prestwich et al. 2007). This line is expected to form in vicinity of hot W-R star 
(high ionization threshold) and trace its orbital motion. However, the accreting compact 
object may ionize significant part of W-R wind in which this line forms. It is then possible 
that the line will form not only in vicinity of W-R star but also in other sectors of binary 
affecting mass estimates of binary components (van Kerkwijk et al. 1996). This is why we 
call the compact object in IC10 X-1 only a ``potentially'' very massive BH.

\section{B. Relevance of the detailed upper limits for advanced LIGO/Virgo}
\label{detUL}

\begin{figure}
\includegraphics[width=1.0\columnwidth]{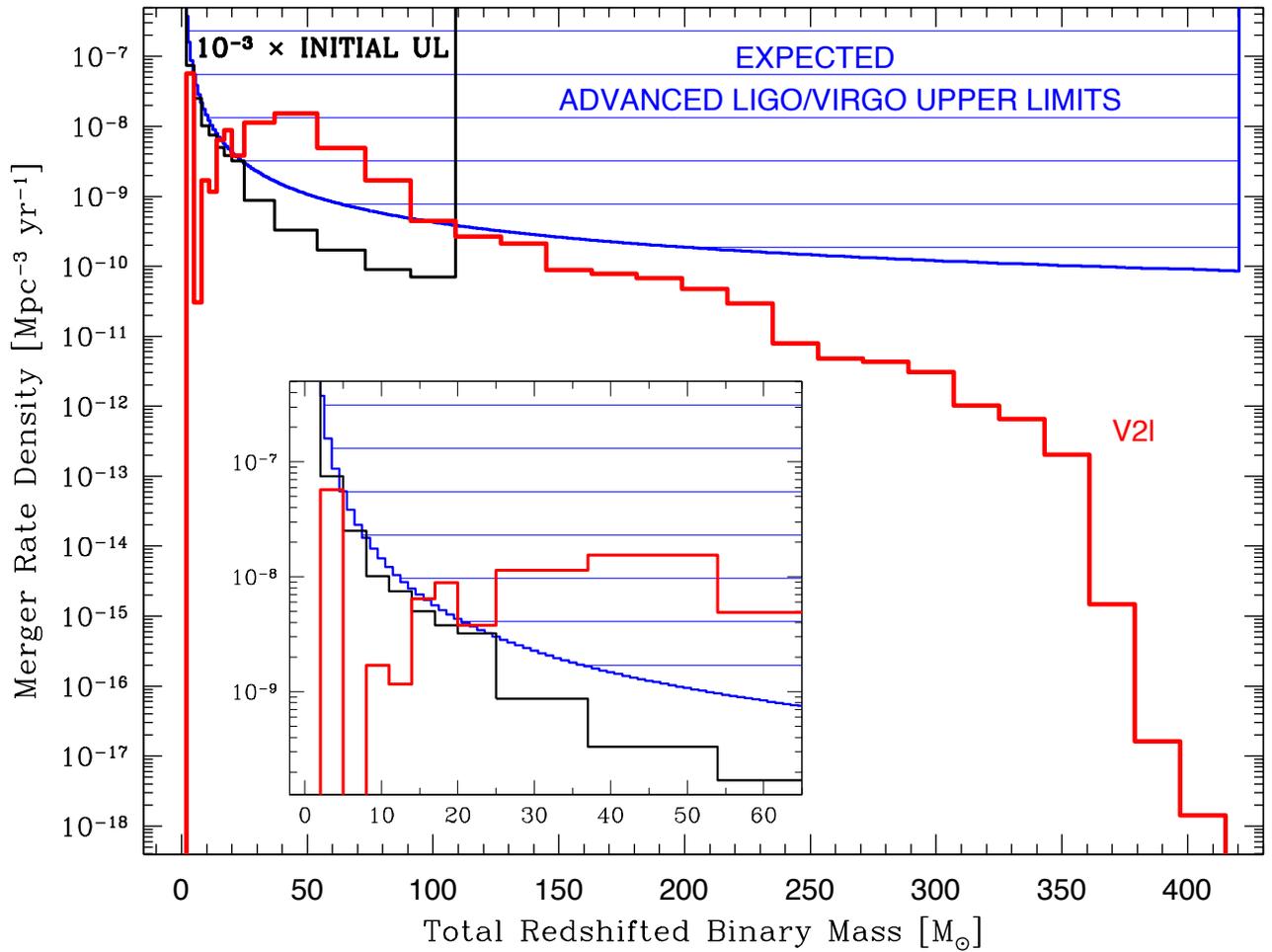}
\vspace*{-5.3cm}
\caption{
Comparison of detailed calculations of the expected advanced LIGO/Virgo upper limits on 
double compact object merger rate density (blue; thick solid line with shaded 
area) with the simple estimate (black; thin solid line). The detailed upper limits are 
obtained with full waveforms (inspiral--merger--ringdown) and with the advanced 
detector sensitivity curve (see Sec.~\ref{aligo}). The simple estimate is calculated 
by multiplying the existing initial LIGO/Virgo upper limits by factor of $10^{-3}$, 
which corresponds to a rough estimate of the difference in volume sampled by initial and advanced 
instruments. Note that above a total merger mass of $25\msun$ the simple estimate 
breaks down (the upper limits are significantly too deep). For comparison we 
show our predicted merger rate density for our standard evolutionary model for low 
metallicity (V2l) and calculated with method II.
}
\label{simple}
\end{figure}

The upper limits for advanced LIGO/Virgo obtained with the detailed calculation (see 
Sec.~\ref{aligo}) can be contrasted with a simple intuitive estimate (e.g., 
Figure~\ref{simple}). For the simple estimate we scale the initial LIGO/Virgo upper limits 
(see Sec.~\ref{iligo}) by a factor of $0.001$, corresponding 
to an increase in the sampled volume by a factor $1000\times$ (and an increase in the horizon 
distance by a factor $10\times$) going from initial to advanced LIGO/Virgo. For low total 
merger mass ($\lesssim25\msun$) both methods give very similar upper limits. However, for 
higher total merger mass ($\gtrsim25\msun$) the simple estimate results in much deeper upper 
limits than expected for advanced LIGO/Virgo. Because advanced LIGO is sensitive to coalescing 
binaries at moderate to significant redshifts, an extrapolation of the initial LIGO upper 
limits must both increase sensitivity ($\times 10^{-3}$ on the $y$ axis) and account for the 
difference between rest-frame and detected total mass (a mass-dependent shift of the $x$ axis).

\begin{deluxetable}{rcccc}
\tablewidth{330pt}
\tablecaption{Merger Rate Density Calculation Methods [Mpc$^{-3}$ yr$^{-1}$] \tablenotemark{a}}
\tablehead{&&Methods&&\\ 
$M_{\rm tot,z}/\msun$\tablenotemark{b} & aLIGO\tablenotemark{c} & I (V2) & II (V2l) & III (V2l)}
\startdata
2-5     & 1.2e-07 & 6.0e-08 & 5.7e-08 & 5.7e-08 \\
5-8     & 3.3e-08 & 2.5e-10 & 3.0e-11 & 1.2e-11 \\
8-11    & 1.6e-08 & 5.1e-09 & 1.7e-09 & 8.7e-10 \\
11-14   & 1.0e-08 & 1.0e-08 & 1.2e-09 & 2.1e-09 \\
14-17   & 6.7e-09 & 2.5e-08 & 6.4e-09 & 2.4e-09 \\ 
17-20   & 4.9e-09 & 5.4e-09 & 8.9e-09 & 6.9e-09 \\ 
20-25   & 3.6e-09 & 1.8e-08 & 3.8e-09 & 6.9e-09 \\ 
25-37   & 2.2e-09 & 2.5e-08 & 1.1e-08 & 6.5e-09 \\ 
37-54   & 1.2e-09 & 1.6e-09 & 1.5e-08 & 1.5e-08 \\ 
54-73   & 7.7e-10 &     0.0 & 4.9e-09 & 6.0e-09 \\ 
73-91   & 5.6e-10 &     0.0 & 1.7e-09 & 4.9e-09 \\ 
91-109  & 4.3e-10 &     0.0 & 4.5e-10 & 1.1e-09 \\
&&&&\\
109-127 & 3.5e-10 &     0.0 & 2.7e-10 & 4.2e-10 \\
127-145 & 2.9e-10 &     0.0 & 2.1e-10 & 3.0e-10 \\
145-163 & 2.5e-10 &     0.0 & 8.8e-11 & 1.9e-10 \\
163-181 & 2.2e-10 &     0.0 & 7.8e-11 & 1.7e-10 \\
181-199 & 2.0e-10 &     0.0 & 6.8e-11 & 1.8e-10 \\
199-217 & 1.8e-10 &     0.0 & 4.7e-11 & 1.7e-10 \\
217-235 & 1.6e-10 &     0.0 & 3.0e-11 & 1.3e-10 \\
235-253 & 1.5e-10 &     0.0 & 7.9e-12 & 6.1e-11 \\
253-271 & 1.4e-10 &     0.0 & 4.8e-12 & 4.5e-11 \\
271-289 & 1.3e-10 &     0.0 & 4.3e-12 & 2.3e-11 \\
289-307 & 1.2e-10 &     0.0 & 3.1e-12 & 2.2e-11 \\
307-325 & 1.1e-10 &     0.0 & 1.0e-12 & 6.9e-12 \\
325-343 & 1.1e-10 &     0.0 & 6.6e-13 & 1.0e-11 \\
343-361 & 1.0e-10 &     0.0 & 2.0e-13 & 3.4e-12 \\
361-379 & 9.8e-11 &     0.0 & 1.5e-15 & 2.0e-13 \\
379-397 & 9.3e-11 &     0.0 & 1.6e-17 & 4.3e-15 \\
397-415 & 9.0e-11 &     0.0 & 1.4e-18 & 5.5e-16 \\
\enddata
\label{t1}
\tablenotetext{a}{
Merger rate density of double compact objects within the horizon of advanced 
LIGO/Virgo obtained with three different methods: I, II, III  for one 
underlying evolutionary model V2 (see Sec.~\ref{calcul} for details).}
\tablenotetext{b}{
Our binning corresponds to the initial LIGO/Virgo low- and high-mass search bins,
extended above $109\msun$ with equal mass bins of $18\msun$.
}
\tablenotetext{c}{Expected advanced LIGO/Virgo upper limits (Sec.~\ref{aligo}).}
\end{deluxetable}

\begin{deluxetable}{rccccc}
\tablewidth{350pt}
\tablecaption{Low-$Z$ Merger Rate Density for initial LIGO/Virgo [Mpc$^{-3}$ yr$^{-1}$] \tablenotemark{a}}
\tablehead{&&&Models&&\\ 
$M_{\rm tot,z}/\msun$\tablenotemark{b} & iLIGO\tablenotemark{c} & V1l & V2l & V3l & V4l }
\startdata
    2--5  & 7.5e-05 & 1.3e-07 & 5.0e-08 & 5.1e-08 & 5.7e-08 \\ 
    5--8  & 2.5e-05 & 9.3e-10 & 5.2e-11 & 1.2e-12 & 7.8e-10 \\ 
   8--11  & 1.0e-05 & 1.6e-08 & 1.7e-09 & 2.9e-10 & 8.6e-10 \\ 
  11--14  & 7.5e-06 & 2.3e-08 & 2.8e-09 & 7.9e-11 & 1.2e-09 \\ 
  14--17  & 5.0e-06 & 9.8e-08 & 1.1e-08 & 1.2e-10 & 9.6e-10 \\ 
  17--20  & 3.8e-06 & 1.6e-08 & 1.6e-09 & 7.9e-12 & 1.2e-09 \\ 
  20--25  & 3.2e-06 & 2.4e-08 & 2.5e-09 & 6.4e-12 & 3.6e-09 \\ 
  25--37  & 8.7e-07 & 4.9e-08 & 1.3e-08 & 4.0e-11 & 1.1e-08 \\ 
  37--54  & 3.3e-07 & 7.9e-09 & 4.5e-09 & 1.4e-11 & 4.1e-09 \\ 
  54--73  & 1.7e-07 & 3.4e-09 & 6.0e-10 & 2.5e-12 & 4.9e-10 \\ 
  73--91  & 9.0e-08 & 8.5e-10 & 2.1e-10 & 7.2e-11 & 2.3e-10 \\ 
 91--109  & 7.0e-08 & 2.8e-11 & 1.6e-11 & 1.5e-11 & 9.2e-12 \\ 
&&&&& \\ 
109--127  &     --- & 4.4e-12 & 3.4e-12 & 7.7e-14 & 1.2e-11 \\ 
127--145  &     --- &     0.0 &     0.0 &     0.0 & 1.2e-15 \\ 
145--163  &     --- &     0.0 &     0.0 &     0.0 &     0.0 \\ 
\enddata
\label{t3}
\tablenotetext{a}{
Merger rate density of double compact objects calculated using full 
inspiral--merger--ringdown waveforms (see Sec.~\ref{method2}). 
Rate  densities are given for the low-metallicity evolution scenario.}
\tablenotetext{b}{
Our binning corresponds to the initial LIGO/Virgo low- and high-mass search bins.   
}
\tablenotetext{c}{
Available initial LIGO/Virgo upper limits for equal mass mergers.
}
\end{deluxetable}

\begin{deluxetable}{rccccc}
\tablewidth{350pt}
\tablecaption{High-$Z$ Merger Rate Density for initial LIGO/Virgo [Mpc$^{-3}$ yr$^{-1}$] \tablenotemark{a}}
\tablehead{&&&Models&&\\ 
$M_{\rm tot,z}/\msun$\tablenotemark{b} & iLIGO\tablenotemark{c} & V1h & V2h & V3h & V4h }
\startdata
    2--5  & 7.5e-05 & 1.6e-07 & 5.7e-08 & 5.8e-08 & 6.8e-08 \\ 
    5--8  & 2.5e-05 & 5.1e-10 & 2.0e-10 & 9.9e-13 & 3.6e-10 \\ 
   8--11  & 1.0e-05 & 1.4e-08 & 2.0e-09 & 3.4e-11 & 3.5e-10 \\ 
  11--14  & 7.5e-06 & 1.1e-08 & 2.5e-09 & 4.3e-11 & 6.1e-10 \\ 
  14--17  & 5.0e-06 & 6.5e-08 & 1.3e-08 & 1.8e-10 & 8.4e-10 \\ 
  17--20  & 3.8e-06 & 6.0e-09 & 7.3e-10 & 4.0e-11 & 6.8e-10 \\ 
  20--25  & 3.2e-06 & 9.2e-09 & 1.7e-09 & 8.9e-12 & 1.7e-09 \\ 
  25--37  & 8.7e-07 & 1.7e-08 & 7.0e-09 & 2.0e-11 & 8.0e-09 \\ 
  37--54  & 3.3e-07 & 3.2e-09 & 2.0e-09 & 6.8e-12 & 1.9e-09 \\ 
  54--73  & 1.7e-07 & 2.0e-09 & 4.3e-10 & 3.1e-12 & 3.1e-10 \\ 
  73--91  & 9.0e-08 & 5.8e-10 & 1.5e-10 & 5.4e-11 & 1.3e-10 \\ 
 91--109  & 7.0e-08 & 1.5e-11 & 1.5e-11 & 1.7e-11 & 1.2e-11 \\ 
&&&&& \\ 
109--127  &     --- & 1.6e-12 & 1.6e-12 &     0.0 & 5.2e-12 \\ 
127--145  &     --- &     0.0 &     0.0 &     0.0 & 6.9e-16 \\ 
145--163  &     --- &     0.0 &     0.0 &     0.0 &     0.0 \\ 
\enddata
\label{t4}
\tablenotetext{a}{
Merger rate density of double compact objects calculated using  full 
inspiral--merger--ringdown waveforms for initial LIGO/Virgo (see Sec.~\ref{method2}).
Rate densities are given for the high-metallicity evolution scenario.}
\tablenotetext{b}{
Our binning corresponds to the initial LIGO/Virgo low- and high-mass search bins.   
}
\tablenotetext{c}{
Available initial LIGO/Virgo upper limits for equal mass mergers.
}
\end{deluxetable}

\begin{deluxetable}{rccccc}
\tablewidth{350pt}
\tablecaption{Low-$Z$ Merger Rate Density for advanced LIGO/Virgo [Mpc$^{-3}$ yr$^{-1}$] \tablenotemark{a}}
\tablehead{&&&Models&&\\ 
$M_{\rm tot,z}/\msun$\tablenotemark{b} & aLIGO\tablenotemark{c} & V1l & V2l & V3l & V4l }
\startdata
    2--5  & 1.2e-07 & 1.5e-07 & 5.7e-08 & 5.4e-08 & 6.2e-08 \\ 
    5--8  & 3.3e-08 & 7.8e-10 & 3.0e-11 & 5.2e-13 & 8.3e-10 \\ 
   8--11  & 1.6e-08 & 1.5e-08 & 1.7e-09 & 6.0e-10 & 7.1e-10 \\ 
  11--14  & 1.0e-08 & 8.8e-09 & 1.2e-09 & 1.6e-10 & 9.7e-10 \\ 
  14--17  & 6.7e-09 & 7.0e-08 & 6.4e-09 & 8.2e-11 & 1.1e-09 \\ 
  17--20  & 4.9e-09 & 9.0e-08 & 8.9e-09 & 1.2e-10 & 1.0e-09 \\ 
  20--25  & 3.6e-09 & 4.3e-08 & 3.8e-09 & 6.3e-11 & 2.7e-09 \\ 
  25--37  & 2.2e-09 & 8.4e-08 & 1.1e-08 & 1.5e-10 & 1.3e-08 \\ 
  37--54  & 1.2e-09 & 7.4e-08 & 1.5e-08 & 1.0e-10 & 1.6e-08 \\ 
  54--73  & 7.7e-10 & 2.3e-08 & 4.9e-09 & 1.1e-10 & 4.9e-09 \\ 
  73--91  & 5.6e-10 & 6.2e-09 & 1.7e-09 & 9.9e-11 & 1.6e-09 \\ 
 91--109  & 4.3e-10 & 4.5e-09 & 4.5e-10 & 3.7e-11 & 4.9e-10 \\ 
&&&&& \\ 
109--127  & 3.5e-10 & 2.5e-09 & 2.7e-10 & 1.7e-11 & 3.8e-10 \\ 
127--145  & 2.9e-10 & 9.3e-10 & 2.1e-10 & 6.0e-12 & 2.5e-10 \\ 
145--163  & 2.5e-10 & 2.7e-10 & 8.8e-11 & 5.0e-12 & 1.2e-10 \\ 
163--181  & 2.2e-10 & 1.5e-10 & 7.8e-11 & 3.6e-12 & 9.9e-11 \\ 
181--199  & 2.0e-10 & 1.1e-10 & 6.8e-11 & 9.3e-13 & 7.1e-11 \\ 
199--217  & 1.8e-10 & 8.1e-11 & 4.7e-11 & 2.8e-12 & 4.5e-11 \\ 
217--235  & 1.6e-10 & 5.4e-11 & 3.0e-11 & 4.9e-12 & 2.9e-11 \\ 
235--253  & 1.5e-10 & 2.4e-11 & 7.9e-12 & 6.0e-12 & 1.8e-11 \\ 
253--271  & 1.4e-10 & 1.0e-11 & 4.8e-12 & 3.0e-13 & 4.8e-12 \\ 
271--289  & 1.3e-10 & 6.4e-12 & 4.3e-12 & 3.7e-14 & 9.9e-13 \\ 
289--307  & 1.2e-10 & 2.8e-12 & 3.1e-12 & 4.3e-14 & 6.9e-13 \\ 
307--325  & 1.1e-10 & 9.0e-13 & 1.0e-12 & 1.7e-14 & 4.9e-13 \\ 
325--343  & 1.1e-10 & 1.0e-12 & 6.6e-13 & 4.0e-15 & 9.0e-13 \\ 
343--361  & 1.0e-10 & 1.8e-13 & 2.0e-13 & 1.0e-18 & 4.6e-13 \\ 
361--379  & 9.8e-11 & 1.8e-15 & 1.5e-15 &     0.0 & 9.7e-14 \\ 
379--397  & 9.3e-11 & 1.2e-17 & 1.6e-17 &     0.0 & 1.3e-16 \\ 
397--415  & 9.0e-11 & 1.9e-18 & 1.4e-18 &     0.0 & 8.2e-19 \\ 
\enddata
\label{t5}
\tablenotetext{a}{
Merger rate density of double compact objects for advanced 
LIGO/Virgo (see Sec.~\ref{method2}). Rate 
densities are given for low-metallicity evolution scenario.}
\tablenotetext{b}{
Our binning corresponds to the initial LIGO/Virgo low- and high-mass search bins 
extended to higher masses.   
}
\tablenotetext{c}{Expected advanced LIGO/Virgo upper limits (Sec.~\ref{aligo}).}
\end{deluxetable}

\begin{deluxetable}{rccccc}
\tablewidth{350pt}
\tablecaption{High-$Z$ Merger Rate Density for advanced LIGO/Virgo [Mpc$^{-3}$ yr$^{-1}$] \tablenotemark{a}}
\tablehead{&&&Models&&\\ 
$M_{\rm tot,z}/\msun$\tablenotemark{b} & aLIGO\tablenotemark{c} & V1h & V2h & V3h & V4h }
\startdata
    2--5  & 1.2e-07 & 1.7e-07 & 6.4e-08 & 5.9e-08 & 6.8e-08 \\ 
    5--8  & 3.3e-08 & 2.7e-10 & 4.5e-11 & 1.7e-13 & 4.6e-10 \\ 
   8--11  & 1.6e-08 & 1.3e-08 & 1.6e-09 & 6.4e-11 & 4.7e-10 \\ 
  11--14  & 1.0e-08 & 4.8e-09 & 6.9e-10 & 9.1e-11 & 5.4e-10 \\ 
  14--17  & 6.7e-09 & 4.4e-08 & 7.2e-09 & 6.9e-11 & 7.1e-10 \\ 
  17--20  & 4.9e-09 & 5.7e-08 & 1.1e-08 & 7.6e-11 & 7.8e-10 \\ 
  20--25  & 3.6e-09 & 2.0e-08 & 3.5e-09 & 3.5e-11 & 1.6e-09 \\ 
  25--37  & 2.2e-09 & 3.3e-08 & 6.5e-09 & 1.0e-10 & 7.5e-09 \\ 
  37--54  & 1.2e-09 & 2.9e-08 & 9.1e-09 & 6.0e-11 & 7.6e-09 \\ 
  54--73  & 7.7e-10 & 7.0e-09 & 2.2e-09 & 7.0e-11 & 2.2e-09 \\ 
  73--91  & 5.6e-10 & 2.9e-09 & 8.0e-10 & 6.7e-11 & 8.4e-10 \\ 
 91--109  & 4.3e-10 & 2.5e-09 & 2.8e-10 & 2.4e-11 & 3.3e-10 \\ 
&&&&& \\ 
109--127  & 3.5e-10 & 1.4e-09 & 1.8e-10 & 1.1e-11 & 2.5e-10 \\ 
127--145  & 2.9e-10 & 5.5e-10 & 1.5e-10 & 4.0e-12 & 1.8e-10 \\ 
145--163  & 2.5e-10 & 1.6e-10 & 5.2e-11 & 2.2e-12 & 7.3e-11 \\ 
163--181  & 2.2e-10 & 7.8e-11 & 4.4e-11 & 3.5e-12 & 7.1e-11 \\ 
181--199  & 2.0e-10 & 6.0e-11 & 4.2e-11 & 2.8e-13 & 4.1e-11 \\ 
199--217  & 1.8e-10 & 4.3e-11 & 2.7e-11 & 1.2e-12 & 2.8e-11 \\ 
217--235  & 1.6e-10 & 3.4e-11 & 1.9e-11 & 2.5e-12 & 1.7e-11 \\ 
235--253  & 1.5e-10 & 1.3e-11 & 5.1e-12 & 4.3e-12 & 1.2e-11 \\ 
253--271  & 1.4e-10 & 5.8e-12 & 2.1e-12 & 3.6e-13 & 3.6e-12 \\ 
271--289  & 1.3e-10 & 3.5e-12 & 2.9e-12 & 4.6e-15 & 4.1e-13 \\ 
289--307  & 1.2e-10 & 1.7e-12 & 1.7e-12 & 1.3e-14 & 2.8e-13 \\ 
307--325  & 1.1e-10 & 7.1e-13 & 5.6e-13 & 1.0e-14 & 2.5e-13 \\ 
325--343  & 1.1e-10 & 4.5e-13 & 4.8e-13 & 3.8e-15 & 2.4e-13 \\ 
343--361  & 1.0e-10 & 1.8e-13 & 1.9e-13 &     0.0 & 5.1e-13 \\ 
361--379  & 9.8e-11 & 1.5e-14 & 4.1e-16 &     0.0 & 4.3e-14 \\ 
379--397  & 9.3e-11 &     0.0 & 1.4e-18 &     0.0 & 1.8e-17 \\ 
397--415  & 9.0e-11 &     0.0 &     0.0 &     0.0 &     0.0 \\ 
\enddata
\label{t6}
\tablenotetext{a}{
Merger rate density of double compact objects for advanced LIGO/Virgo 
(see Sec.~\ref{method2}). Rate densities are given for the high-metallicity 
evolution scenario.}
\tablenotetext{b}{
Our binning corresponds to the initial LIGO/Virgo low- and high-mass search bins 
extended to higher masses.   
}
\tablenotetext{c}{Expected advanced LIGO/Virgo upper limits (Sec.~\ref{aligo}).}
\end{deluxetable}

\begin{figure}
\includegraphics[width=1.0\columnwidth]{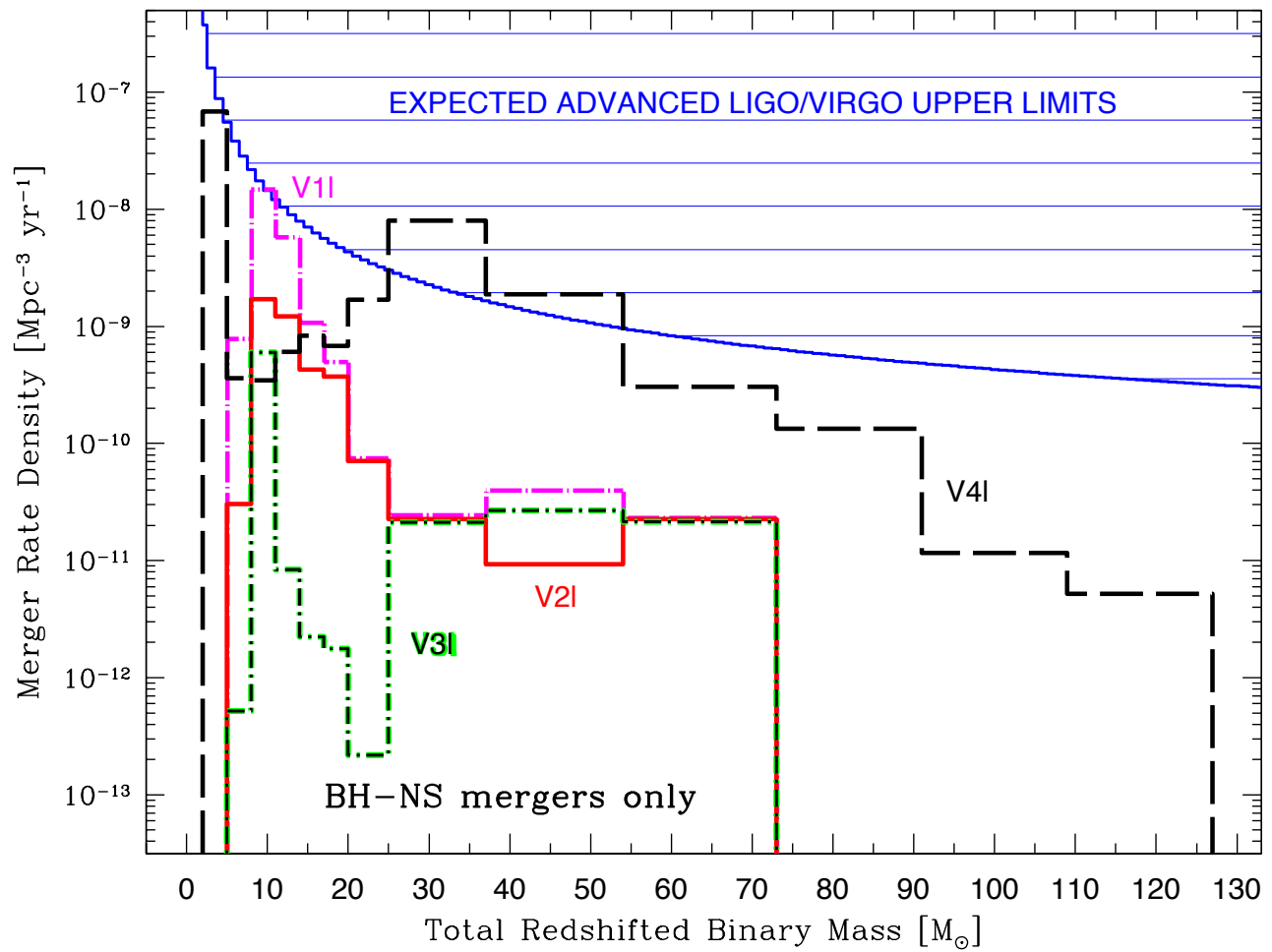}
\vspace*{-5.3cm}
\caption{
Merger rate density for BH-NS systems for our low metallicity evolution  
scenario predicted (with method II) for advanced LIGO/Virgo. Note that only
the delayed SN explosion model (V4) and the optimistic model (V1) make 
detections likely, while the other models indicate non-detection even at 
the full advanced LIGO/Virgo sensitivity.  
}
\label{f6}
\end{figure}

\end{document}